\documentclass{jfm}

\usepackage{graphicx}
\usepackage{newtxtext}
\usepackage{newtxmath}
\usepackage{natbib}
\usepackage{hyperref}
\usepackage{multirow} 
\usepackage[percent]{overpic} 
\usepackage{tikz}
\usetikzlibrary{arrows.meta}
\usepackage{booktabs}
\usetikzlibrary{calc}

\hypersetup{
    colorlinks = true,
    urlcolor   = blue,
    citecolor  = black,
}

\title{Uncovering bistability phenomena in two-layer Couette flow experiments using nonlocal evolution equations}

\author{X. Wang\aff{1},
P. Germain\aff{1}
\and
  D. T. Papageorgiou\aff{1}  \corresp{\email{d.papageorgiou@imperial.ac.uk}}
 }

\affiliation{\aff{1}Department of Mathematics, Imperial College London, London SW7 2AZ, UK}

\begin{document}
\maketitle

\begin{abstract}
This paper investigates the stability of interfacial long waves in two-layer plane Couette flow using a nonlinear, nonlocal asymptotic model derived from the Navier-Stokes equations and valid for thin upper layers. Nonlocality enters through a coupling of the thin and main layers, and crucial inertial effects are retained. The models generically support bistability phenomena observed in experiments where two stable travelling waves, one unimodal and the other bimodal, are recorded at the same lid velocity. In direct comparisons with experiments, the models show remarkable agreement, both qualitatively and quantitatively. The two stable travelling waves are identified and their basins of attraction characterised via large-time computations for different initial conditions. We also identify a new symmetry-breaking travelling-wave branch bifurcating from the bimodal family, compute higher-wavenumber travelling-wave branches, and present time-periodic orbits arising via Hopf bifurcation.


\end{abstract}

\section{Introduction}
\label{sec:headings}

Two-fluid plane Couette flows were first shown by \citet{Yih1967} to be unstable to long waves at nonzero Reynolds numbers. 
The instability depends on the viscosity and thickness ratios of the two layers. Subsequent analyses established the ``thin-layer effect"---namely that if the thin layer is more viscous, the flow is linearly unstable \citep{HooperBoyd1983,Hooper1985,Renardy1985,Renardy1987}. Weakly nonlinear reductions connected these results to dispersive Kuramoto--Sivashinsky-type amplitude equations \citep{HooperGrimshaw1985,HooperGrimshaw1988,CharruFabre1994}.
We are motivated by certain key experimental observations of \cite{BartheletCharruFabre1995}, who studied wave formation in a two-fluid annular Couette device driven by a rotating upper lid. The depth of the device's annulus is smaller than its width, and so a two-dimensional assumption is reasonable, at least away from end walls.
The geometry of the experiment is ideal for comparisons to certain nonlinear reduced dimension evolution equation models for several reasons: (i) observed waves are periodic with a period at most equal to the perimeter, $L_w$ say, of the annulus, (ii) $L_w$ unambiguously sets the period to be used in model equations, and (iii) any bifurcations from unimodal, i.e. $L_w$-periodic, to bimodal $\frac{1}{2}L_w$-periodic waves arise naturally and are within reach of analytical descriptions. Indeed, \cite{BartheletCharruFabre1995} reported long–wave asymmetry and, at higher upper-lid speeds, the existence of bistability, with unimodal and bimodal travelling waves coexisting.
Recent modelling developments have introduced an inertia–retaining nonlocal model for two–layer Couette flow, in which the thick lower layer enters via an Orr--Sommerfeld problem at leading order \citep{KalogirouPapageorgiou2016, KalogirouCimpeanuKeavenyPapageorgiou2016R1}, enabling quantitative comparisons at experimentally relevant Reynolds numbers. 

We also note closely related mathematical studies that use the same travelling wave machinery in analogous settings: \citet{PapageorgiouTanveer2019} developed and analysed a nonlocal model when the thin layer is the lower one; \citet{KatsiavriaPapageorgiou2022} incorporated interfacial slip in multilayer shear flows and examined its impact on nonlinear wave selection; and \citet{PapageorgiouTanveer2023} demonstrated that even modest interfacial slip can have a singular effect in otherwise stable two-layer shear flows. This finding was shown to be quite generic in the linear stability study of the full problem by \citet{Katsiavria_Papageorgiou_2024}. In the presence of pressure-driven flow, analogous techniques can be used to derive and study model equations and compare results with simulations---see \citet{Kalogirou_Cimpeanu_Blyth_2020, Kalogirou_Blyth_2023}.


The models used in this study are asymptotically valid for a thin, more viscous upper layer lying over a thick, denser layer, e.g., mineral oil/glycerine-water systems used in the experiments. Hence, this study focuses on reproducing experimental cases with thickness ratio \(d=0.25\),  where $d$ is the ratio of the mean upper layer thickness to that of the lower one. The value $d=0.25$ is the smallest one reported in the experiments, and for direct comparison we carried out model computations for the different lid velocities reported by \cite{BartheletCharruFabre1995}, in particular cases which exhibit bistability (e.g. figure~22 in \cite{BartheletCharruFabre1995}). 

We stress the novelty of the present work by comparing it with \citet{KalogirouCimpeanuKeavenyPapageorgiou2016R1}, which, to our knowledge, provided the first detailed joint study of the current nonlocal model with direct numerical simulations and experimental measurements for two-layer Couette flow. First, while \citet{KalogirouCimpeanuKeavenyPapageorgiou2016R1} focused primarily on time-dependent comparisons between DNS, model simulations and experiments, explicit bifurcation and stability diagrams for the unimodal and bimodal travelling waves were not presented; here we compute and continue these travelling-wave branches and their stability boundaries systematically (see \S \ref{sec:fats}). Second, \citet{KalogirouCimpeanuKeavenyPapageorgiou2016R1} reported unsatisfactory quantitative comparisons in the bistability regime; in the present paper this discrepancy is resolved via improved agreement with the experimental data and, furthermore, the basins of attraction of the two coexisting stable travelling waves are separated (see \S \ref{sec:comparison}). The reason for this discrepancy is due to the use of a non-local model equation in 
\citet{KalogirouCimpeanuKeavenyPapageorgiou2016R1} that is appropriate for a lower thin layer rather than an upper one used here, as derived and studied by \citep{KatsiavriaPapageorgiou2022,Katsiavria_Papageorgiou_2024,PapageorgiouTanveer2023}. Third, we identify a new symmetry-breaking travelling-wave branch that bifurcates from the $\pi$-periodic bimodal branch and persists as a local, and over parts of parameter space global, attractor over a broad range of $\Lambda$ (see \S \ref{sec:newbranch}). Fourth, we map the higher-wavenumber branches, and report periodic orbits arising via Hopf bifurcations (see \S\ref{sec:overview}). These higher-wavenumber families, the symmetry-breaking branch, and the periodic dynamics were not reported in \citet{KalogirouCimpeanuKeavenyPapageorgiou2016R1} nor in \citet{BartheletCharruFabre1995}. Finally, we broaden the scope of experimental validation at $d=0.25$ beyond the three (distinct Reynolds number) cases in figure~3 of \citet{KalogirouCimpeanuKeavenyPapageorgiou2016R1}: in \S \ref{sec:IP} we report the interfacial profiles of time-dependent model comparisons for five cases and in \S \ref{sec:HA} harmonic-amplitude diagnostics for four cases, thereby covering essentially the full set of available $d=0.25$ measurements reported by \citet{BartheletCharruFabre1995}.

\section{The experiments of \citet{BartheletCharruFabre1995}}\label{sec:barthelet1995}

The device of \citet{BartheletCharruFabre1995} is given in their figure~2. The upper lid is driven at a prescribed speed \(U\), and the interfacial amplitude is monitored at a fixed azimuthal location. A lighter and more viscous layer lies above a heavier less viscous one, with thickness ratio \(d=h_2/h_1\) (upper to lower layer thickness ratio). For \(d=0.25\), the smallest value examined, the experiments report a supercritical onset of travelling waves whose wavelength equals the device circumference; at higher speeds a bistable response is observed, in which unimodal (\(L_w\)-periodic) and bimodal (\(\frac{1}{2}L_w\)-periodic) waves coexist (they are excited by different initial conditions). Fluid properties for the pair labelled “\(1\mathrm{d}\)–2” in the experiments are collected in table~\ref{tab:exp-phys}. Since our model is derived under the assumption $d \ll 1$, other experiments having $d = 0.33, 0.42$ and $0.82$, are not considered here. Although $d = 0.25$ is not too small, the model predicts many of the experimental observations with remarkable agreement. We expect even better agreement for smaller $d$ but we have not found such cases in the literature.

\begin{table}
  \centering
  \begin{tabular}{lc@{\qquad}lc}
    Physical parameter & Value & Physical parameter & Value \\  
    Channel mean diameter \(D\) & \(0.40\ \mathrm{m}\) & Channel width \(W\) & \(0.04\ \mathrm{m}\) \\
    Channel height (depth) \(d_{\mathrm{phys}}\) & \(0.02\ \mathrm{m}\) & Thickness ratio \(d=h_2/h_1\) & \(0.25\) \\
    Lower fluid viscosity \(\mu_1\) & \(0.0108\ \mathrm{Pa\,s}\) & Upper fluid viscosity \(\mu_2\) & \(0.0297\ \mathrm{Pa\,s}\) \\
    Lower fluid density \(\rho_1\) & \(1142\ \mathrm{kg\,m^{-3}}\) & Upper fluid density \(\rho_2\) & \(846\ \mathrm{kg\,m^{-3}}\) \\
    Viscosity ratio \(m=\mu_2/\mu_1\) & \(2.76\) & Density ratio \(r=\rho_2/\rho_1\) & \(0.741\) \\
    Surface tension \(\gamma\) & \(0.03\ \mathrm{Pa\,m}\) &  &  \\
  \end{tabular}
    \caption{Physical parameters from the experiments of \citet{BartheletCharruFabre1995} (fluid pair \(1\mathrm{d}\)–2) used in the comparisons (as in \citet{KalogirouCimpeanuKeavenyPapageorgiou2016R1}).}
    \label{tab:exp-phys}
\end{table}

The experiments at \(d=0.25\) identify a critical upper-plate speed, above which a supercritical bifurcation produces a travelling wave with wavelength equal to \(L_w=\pi D=1.257\,\mathrm{m}\), implying a dimensionless period $2L = L_w/d_{\mathrm{phys}} \approx 63$---see table~\ref{tab:exp-phys} for the definition of $d_{\mathrm{phys}}$ (ultimately we scale to $2\pi$-periodic domains). The interfacial amplitude is recorded by a fixed probe, and spatial profiles constructed by using the measured wave speed. The dimensionless undisturbed interfacial speed is
\begin{align}
U_s \;=\; 1 - \frac{\varepsilon}{\,m - (m-1)\varepsilon\,},\qquad \varepsilon = \frac{h_1}{h_1+h_2} = \frac{d}{1+d}.\label{eq:Us}
\end{align}
In the model computations we record the interface at \(x=0\), where \(x := x_{\mathrm{lab}} - U_s t\) is the coordinate moving with the undisturbed interface. A \(2\pi\)-periodic travelling wave \(H(x-Ct)\) has temporal period \(2\pi/C\) in this frame. The experiments measure amplitudes at a fixed laboratory position \(x_{\mathrm{lab}} = 0\), and hence the model period must be multiplied by the factor \(C/U_s\). We scale the oscillatory part as in the experiments by defining \(H(t)=y(0,t)-(1-\varepsilon)\) and compute \(A_+=\max_t H(t)\) and \(A_-=\min_t H(t)\). The scaled amplitude is \(H(t)/H_{\mathrm{sat}}\), where \(H_{\mathrm{sat}}=\tfrac12\big(A_+ - A_-\big)\). The model also assumes $\rho_1 = \rho_2 = \rho$ (in the experiment they differ somewhat, see table \ref{tab:exp-phys}).  The resulting non-dimensionalised parameters are
\begin{equation}
  Re_1=\frac{\rho U d_{\mathrm{phys}}}{\mu_1},\qquad
  Re_2=\frac{\rho U d_{\mathrm{phys}}}{\mu_2},\qquad
  We=\frac{\rho U^2 d_{\mathrm{phys}}}{\gamma},\qquad
  Ca=\frac{\mu_2 U}{\gamma}.
  \label{eq:dimless-groups}
\end{equation}

\section{Mathematical model}
\label{sec:model}

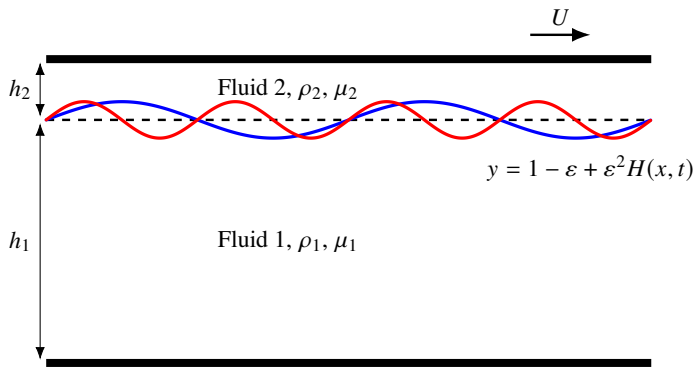
\begin{figure}
\centering
\begin{tikzpicture}[x=8cm,y=5cm] 

  \draw[line width=3pt] (0,0) -- (1,0);
  \draw[line width=3pt] (0,1) -- (1,1);

  \draw[->, >=Latex, line width=0.9pt] (0.8,1.08) -- (0.90,1.08)
       node[midway, above] {$U$};

  \draw[dashed, line width=0.9pt] (0,0.8) -- (1,0.8);

\node[below=10pt] at (0.9,0.8) {$y = 1 - \varepsilon + \varepsilon^{2} H(x,t)$};    

\draw[blue, line width=1.2pt, line cap=round, smooth, samples=200, domain=0:1, variable=\x]
  plot ({\x},{0.8 + 0.06*sin(4*pi*\x r)});

\draw[red, line width=1.2pt, line cap=round, smooth, samples=200, domain=0:1, variable=\x]
  plot ({\x},{0.8 + 0.06*sin(8*pi*\x r)});

  \draw[<->, >=Latex] (-0.01,0.81) -- (-0.01,0.99)
       node[midway, left] {$h_2$};
  \draw[<->, >=Latex] (-0.01,0.01) -- (-0.01,0.79)
       node[midway, left] {$h_1$};

  \node at (0.4,0.9) {Fluid 2, $\rho_2$, $\mu_2$};
  \node at (0.4,0.40) {Fluid 1, $\rho_1$, $\mu_1$};

\end{tikzpicture}
\caption{Two superposed fluid layers in a channel, driven by the upper-plate shear with speed $U$. Blue and red curves represent unimodal and bimodal respectively. $d = h_2/h_1 = 0.25$ and $U_L = 0.138\,\mathrm{m\,s^{-1}}$ fixed across the cases, while $U/U_L$ varies.}
\label{fig:Geometry}
\end{figure}

Following \citep{KalogirouPapageorgiou2016, KalogirouCimpeanuKeavenyPapageorgiou2016R1}, the derivation starts from the Navier--Stokes equations, assumes a thin upper film, and develops asymptotic solutions that yield a nonlinear, nonlocal evolution equation for the interface; all other fields follow directly from this.
Lengths are non-dimensionalised with \(d_{\mathrm{phys}}\), velocities with the upper-plate speed \(U\), time with $d_{\mathrm{phys}}/U$, and pressures with $\rho U^2$. The parameter $\varepsilon$ defined in \eqref{eq:Us} is taken to be small. The interface is at
$y = 1 - \varepsilon + \varepsilon^{2} H(x,t)$, see figure~\ref{fig:Geometry}. An evolution equation for $H$ follows by solving in the thin upper layer where a lubrication solution holds, and matching with the thick lower layer governed by the linearised Navier-Stokes equations, leading naturally to a nonlocal
inertial response. At the interface we enforce the kinematic condition, continuity of velocities and stresses. The resulting dimensionless equation holds on $2L$-periodic domains.
Normalising to $2\pi$-periodic solutions using 
\((t,x,H)\to (t/\nu,\,x/\sqrt{\nu},\,\sqrt{\nu}\,H)\), where $\nu = (\pi/L)^2 = 0.01$, the latter value dictated by the experiments, yields 
\begin{align}
\label{eq:main}
H_t + H\,H_x + \nu\,H_{xxxx} 
  + \sum_{k\in\mathbb{Z}} \mathcal{N}[k]\,\hat H_k\,e^{ikx} = 0,\\
\label{eq:N-def}
\mathcal{N}[k] = -\,i\,\frac{\Lambda}{2\nu}\,k\,\sqrt{\nu}\,F''\!\left(1;\,\alpha\right),
\qquad \alpha = k\,\sqrt{\nu}.
\end{align}
Here $\Lambda = 3Ca_0/2m$ with $Ca_0 = Ca/\varepsilon = O(1)$, so that surface tension is retained. 
The function \(F(y;\alpha)\) is determined by a quasi-static inhomogeneous Orr–Sommerfeld problem,
\begin{equation}\label{eq:Orr-S}
\Bigl(\tfrac{d^2}{dy^2}-\alpha^2\Bigr)^{\!2} F 
\;-\; i\,\alpha\,R\,(y-1)\,\Bigl(\tfrac{d^2}{dy^2}-\alpha^2\Bigr) F \;=\; 0,
\quad 0<y<1,
\end{equation}
where $R = Re_1$ is the Reynolds number of the thick (lower) layer. We have
no-slip at the lower wall $y=0$, and interfacial conditions at $y=1$,
\begin{equation}\label{eq:BCs}
F(0)=F'(0)=0, 
\qquad 
F(1)=0, 
\qquad 
F'(1)=\frac{m-1}{m}.
\end{equation}
The quantity \(F''(1;\alpha)\) in equation~\eqref{eq:N-def} can be obtained numerically from \eqref{eq:Orr-S}-\eqref{eq:BCs}, or in terms of Airy functions as given in \citet[§3]{PapageorgiouTanveer2023}. 

The parameters in the model are $R$, $\Lambda$, $m$ and $\nu$. For a given experiment, $m$, $\nu$ are fixed, and $R$ is proportional to $U$. Although we define earlier \(\Lambda=3\,Ca_0/(2m)\), in practice, \(Ca=\mu_2U/\gamma\) is known from the experimental data, but \(Ca_0\) is not independently measurable (it is defined through the canonical scaling \(Ca=\varepsilon Ca_0\)), so for quantitative comparisons at finite \(\varepsilon\) we treat \(\Lambda\) as an adjustable parameter that is consistent with \(Ca\) being small.

\section{In-depth study of the bistability case}\label{sec:bistability}

In this section, we compute travelling wave solutions of \eqref{eq:main}, identify their stability, study their dynamics, and compare with available experiments. We also delineate the basins of attraction of unimodal (branch 1) and bimodal (branch 2) travelling waves. Moreover, we focus on a symmetry-breaking travelling-wave family (branch~2$^\ast$) that bifurcates from branch~2 and yields further multistability. Finally, we map the broader set of travelling-wave families supported by the model, including higher-wavenumber branches (branches~3 and~4), and describe time-periodic attractors arising through Hopf bifurcations. Guided by the experiments (fluid pair 1d-2, see table \ref{tab:exp-phys}), we set $R = 709$, $\nu = 0.01$, $m = 2.76$, throughout.

\subsection{Finite-amplitude travelling waves and their stability}\label{sec:fats}

We seek $2\pi$–periodic travelling waves of \eqref{eq:main} of the form $H(x,t)=H(z)$ where $z=x-Ct$ and $C$ is the wave speed to find 
\begin{equation}\label{eq:tw-ode}
  -C\,H_z + H\,H_z + \nu\,H_{zzzz} + \sum_{k\in\mathbb{Z}} \mathcal{N}[k]\,\hat H_k\,e^{ikz}=0.
\end{equation}
We represent $H$ by a truncated Fourier series $H(z)=\sum_{k=-N}^{N}\widehat H_k\,e^{ikz}$,
where $\widehat H_{-k}=\overline{\widehat H_k}$, overlines denote complex conjugation, and $\widehat H_{0}=0$, i.e. $\int_0^{2\pi}H(z)dz=0$. 
Using  $\mathcal{N}[-k]=\overline{\mathcal{N}[k]}$ and $\mathcal{N}[0]=0$, the Fourier coefficients of \eqref{eq:tw-ode} satisfy
\begin{equation}\label{eq:fourier-nonlinear}
  -ik\,C\,\widehat H_k \;+\; \frac{ik}{2}\,(\widehat H*\widehat H)(k) \;+\; \nu\,k^4\,\widehat H_k \;+\; \mathcal{N}[k]\,\widehat H_k \;=\; 0,\quad k \in \mathbb{Z}^{+},
\end{equation}
where $(\widehat{H}\ast \widehat{H})(k) \;=\; \sum_{j\in\mathbb{Z}} \widehat{H}_j\,\widehat{H}_{k-j}$. It is sufficient to consider non-negative $k$ since $H(z)$ is real.
Translation invariance is removed by a phase condition; we impose
\begin{equation}\label{eq:phase}
  \Im\,\widehat H_{k_0}=0.
\end{equation}
For branch 1 solutions we set \(k_{0}=1\) and for branch 2, for which \(\widehat{H}_{k}=0\) for all \(k\) odd, we set \(k_{0}=2\) instead. Equations \eqref{eq:fourier-nonlinear}–\eqref{eq:phase}, define a nonlinear algebraic system for the $2N$ real unknowns $\Big(\Re\,\widehat H_1,\,C,\,\Re\,\widehat H_2,\,\Im\,\widehat H_2,\,\ldots,\,\Re\,\widehat H_N,\,\Im\,\widehat H_N\Big)$, which we solve by Newton iteration---see \citet{PapageorgiouTanveer2023} also.

To determine linear stability of computed travelling waves, we add perturbations $\delta\widehat U_k e^{\sigma t}$ to $\widehat H_k$ and linearise with respect to $\delta$ to find the eigenvalue problem
\begin{equation}\label{eq:linstab}
  -\big(\nu\,k^4 + \mathcal{N}[k] - i C k\big)\,\widehat U_k \;-\; ik\,(\widehat U*\widehat H)(k) \;=\; \sigma\,\widehat U_k,
  \quad k\in\{-N,\ldots,-1,1,\ldots,N\}.
\end{equation}
This leads to finding the eigenvalues of a $2N\times2N$ complex-valued matrix with eigenvector $(\widehat U_1,\widehat U_{-1},\widehat U_2,\widehat U_{-2},\ldots,\widehat U_N,\widehat U_{-N})^T$.

\begin{figure}
  \centering

  \begin{minipage}{0.40\linewidth}
    \begin{overpic}[width=\linewidth]{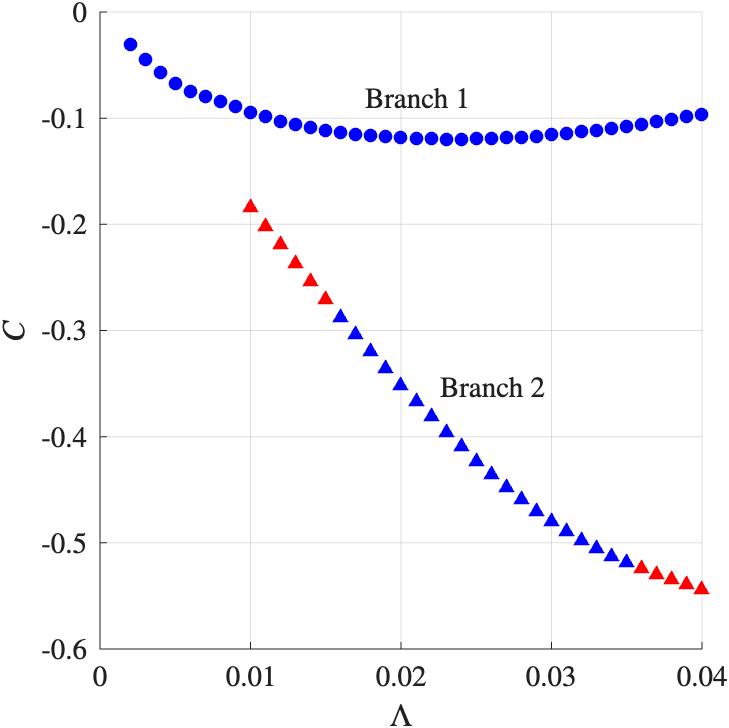} 
      \put(0,99){(\textit{a})}
    \end{overpic}
  \end{minipage}\hspace{0.02\textwidth}
  \begin{minipage}{0.40\linewidth}
    \begin{overpic}[width=\linewidth]{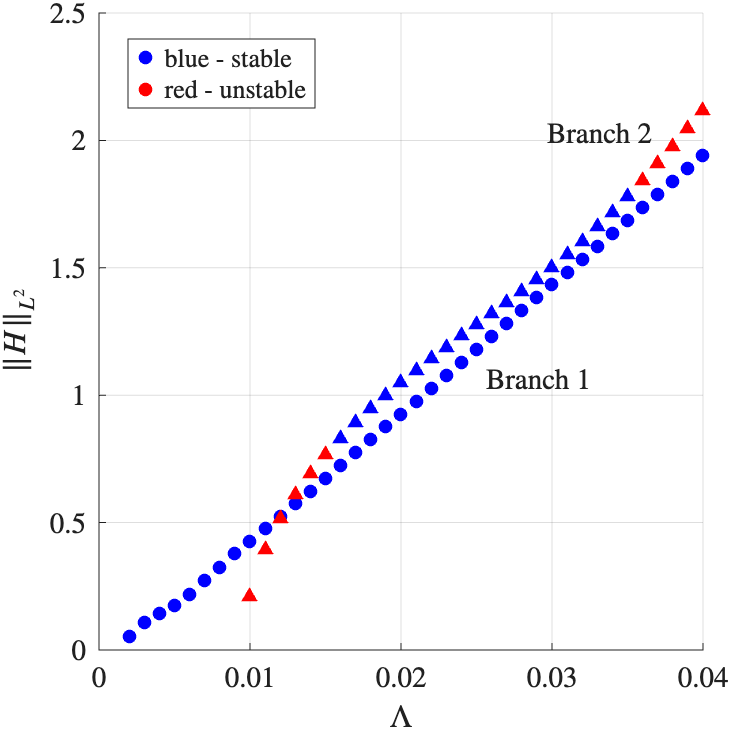} 
      \put(0,99){(\textit{b})}
    \end{overpic}
  \end{minipage}
  
  \begin{minipage}{0.40\linewidth}
    \begin{overpic}[width=\linewidth]{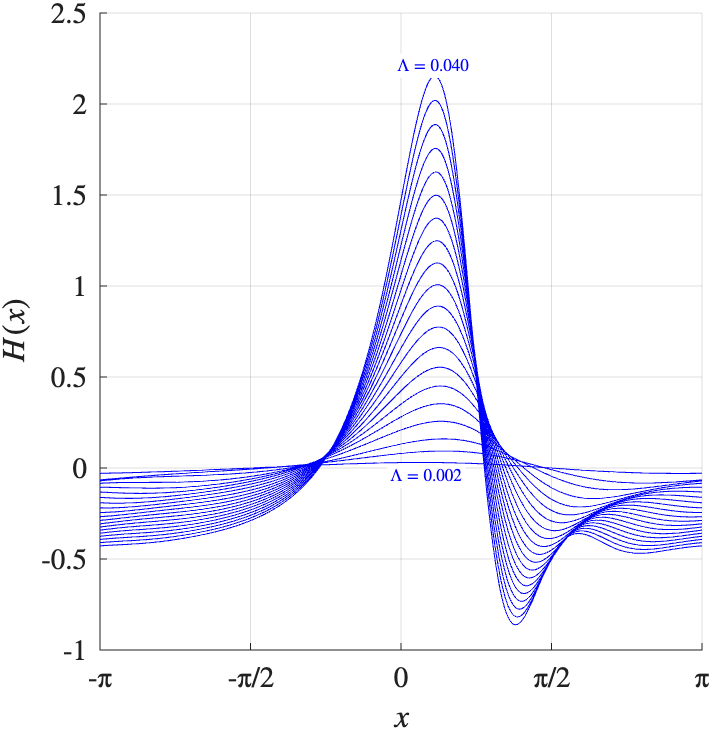} 
      \put(0,99){(\textit{c})}
    \end{overpic}
  \end{minipage}\hspace{0.02\textwidth}
  \begin{minipage}{0.40\linewidth}
    \begin{overpic}[width=\linewidth]{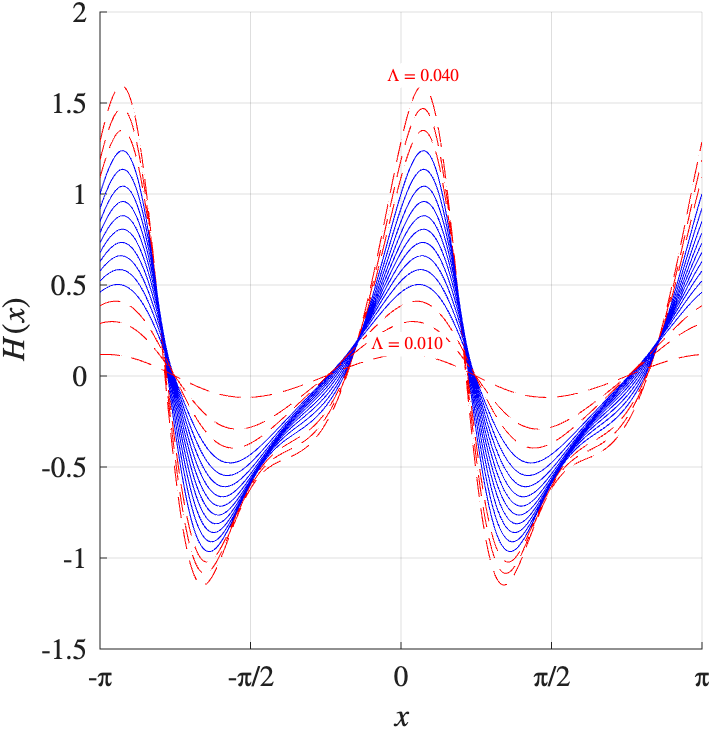} 
      \put(0,99){(\textit{d})}
    \end{overpic}
  \end{minipage}
  \caption{Bifurcation diagrams of branch~1 and branch~2, computed travelling waves and their stability for $R = 709$, $\nu = 0.01$, $m = 2.76$.\quad (\textit{a}) Wave speed $C$ versus $\Lambda$;\quad (\textit{b}) $L^2$-norm versus $\Lambda$.\quad (\textit{c}) Wave profiles for branch 1.\quad (\textit{d}) Wave profiles for branch 2.}
  \label{fig:bistability}
\end{figure}

Figure~\ref{fig:bistability} presents results for both branches along with their stability---blue curves/symbols are used for stable states and red for unstable ones. The upper row shows the bifurcation branches in the $\Lambda-C$ and $\Lambda-\| H\|_{L^2}$ space. We observe that the fundamental \(2\pi\)-periodic family (branch 1) is stable just beyond onset and remains stable up to \(\Lambda\approx 0.040\). At this value the dominant eigenvalue of \eqref{eq:linstab} crosses the imaginary axis with nonzero frequency, and branch 1 loses stability via a Hopf bifurcation. The $\pi$-periodic family (branch 2) emerges at $\Lambda\approx 0.010$ via a pitchfork bifurcation from the flat state and is initially unstable; it stabilises at \(\Lambda\approx 0.016\) and remains stable until \(\Lambda\approx 0.036\). There is a clear bistability window \(0.016\lesssim \Lambda \lesssim 0.036\) in which branches 1 and 2 coexist and are linearly stable, explaining the phenomena observed in experiments.

Indeed, at \(\Lambda\approx 0.036\), branch~2 undergoes a pitchfork bifurcation with a real eigenvalue, giving rise to a new steady branch. The new branch has norm very close to that of branch 2 and so is not depicted here, instead we will discuss this new branch in details in \S ~\ref{sec:newbranch}.

\subsection{Comparison with the experiments of \citet{BartheletCharruFabre1995}}\label{sec:comparison}

For time-dependent solutions, we integrate \eqref{eq:main} numerically using a Fourier spectral discretisation in space and a fourth-order exponential-time-differencing Runge--Kutta method in time.
The experiments for \(d=0.25\) report a bistable window at the highest plate speeds, in which unimodal ($2L$-periodic) and bimodal ($L$-periodic) waves coexist and are achieved by modifying the initial data \citep{BartheletCharruFabre1995}. We target the competing attractors by using sinusoidal small-amplitude initial conditions $H(x,0)=a\sin(x)$, $H(x,0)=a\sin(2x)$, for branches 1 and 2, respectively. We typically set  $a = 10^{-3}$ and fix $\Lambda = 0.020$ to reproduce the observed bistability. The first initial condition saturates to a \(2L\)-periodic wave with pronounced front–trough asymmetry, while the second remains bimodal throughout the evolution, with only even harmonics present.

\begin{figure}
  \centering
  \begin{minipage}{0.59\linewidth}
    \centering
    \begin{overpic}[width=\linewidth]{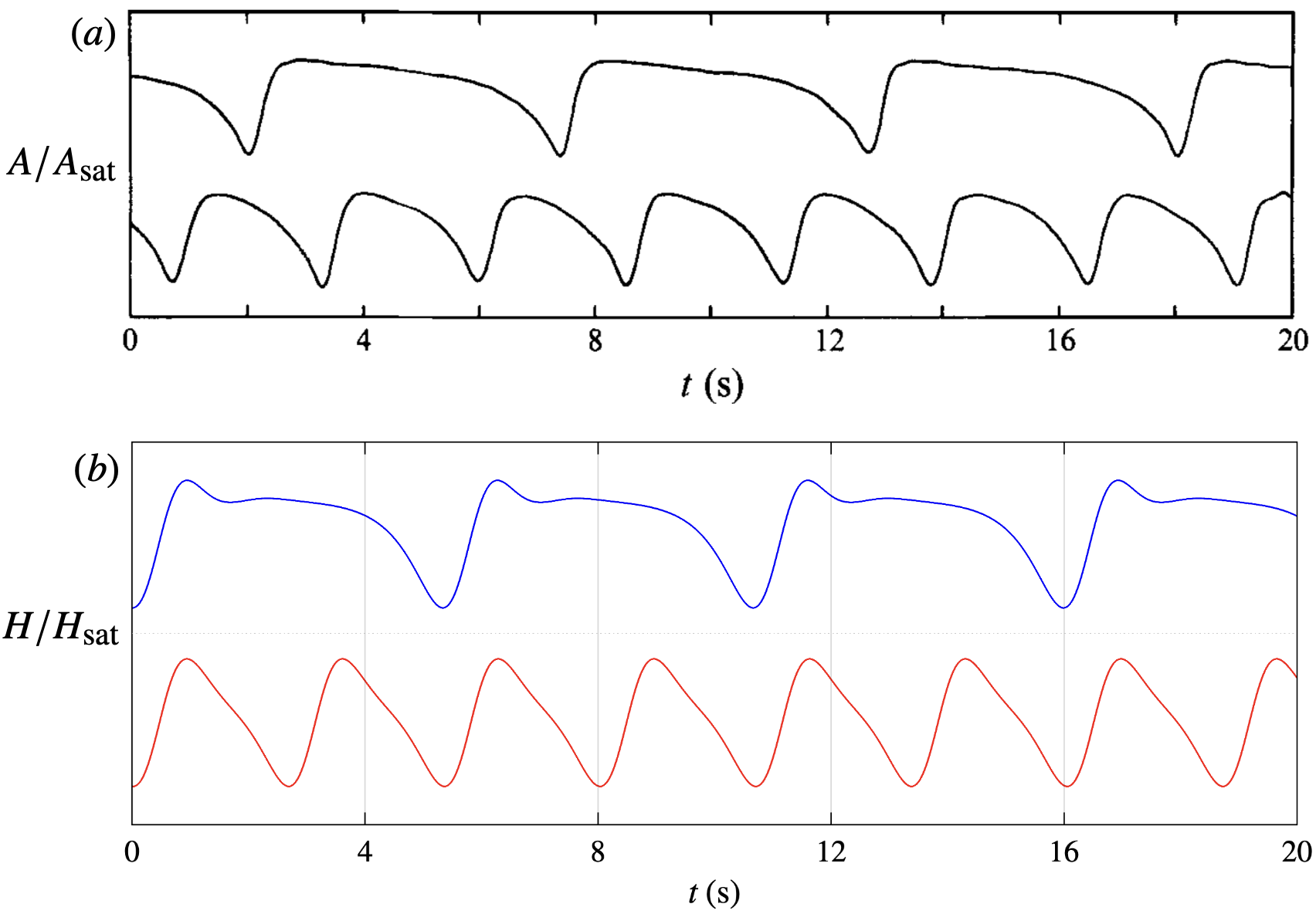}
    \end{overpic}
  \end{minipage}\hfill
  \begin{minipage}{0.39\linewidth}
    \centering
    \begin{overpic}[width=\linewidth]{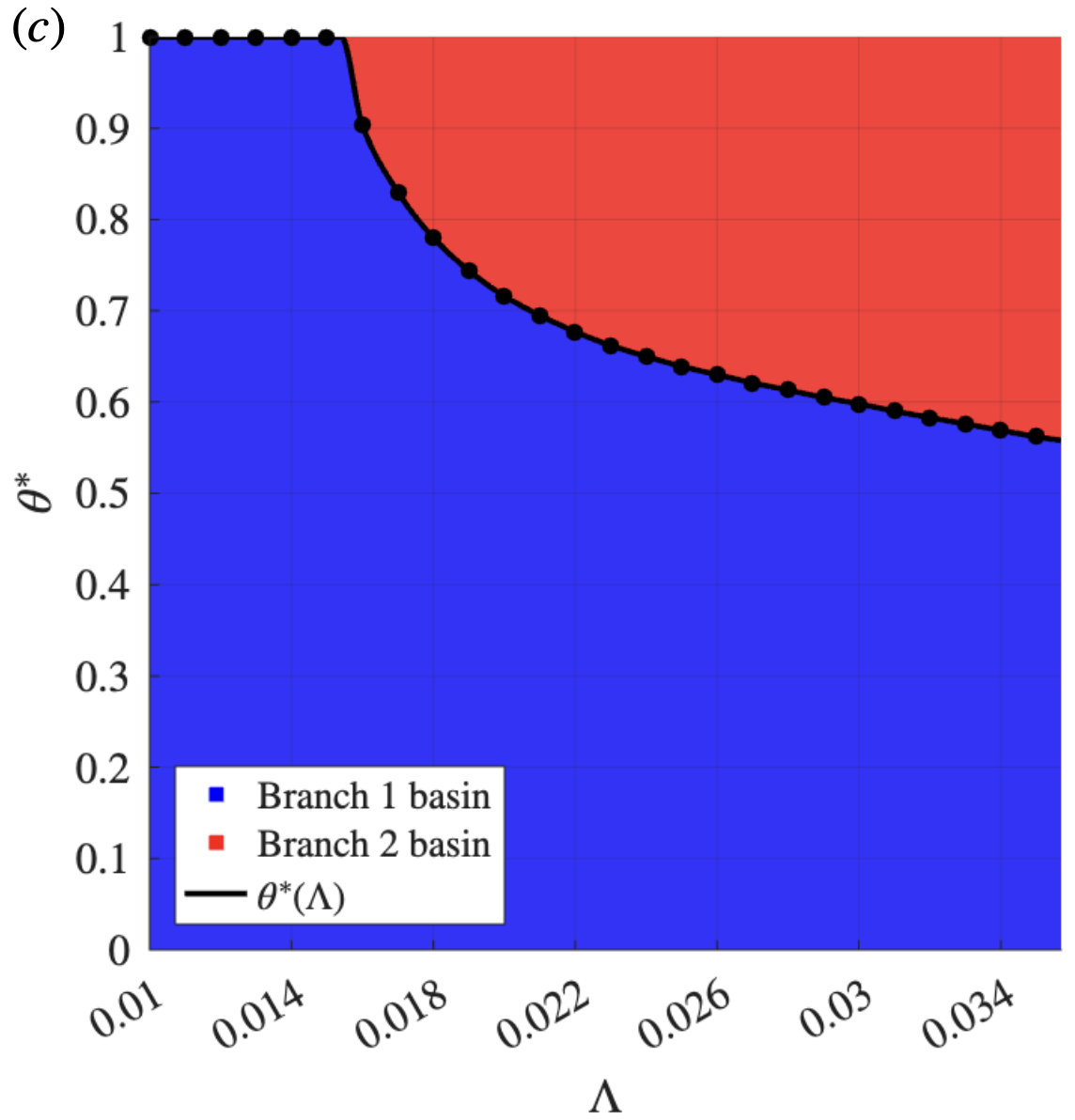}
    \end{overpic}
  \end{minipage}

  \caption{Evolution of the interfacial position: dominated by the fundamental and by the second harmonic.\quad (\textit{a}) Experimental traces (from \citet{BartheletCharruFabre1995}, p.~49);\quad (\textit{b}) model computation;\quad (\textit{c}) basins of attraction; bistability for $\Lambda \in [0.0155, 0.0357]$. }
  \label{fig:main}
\end{figure}

A direct comparison to the experiment is given in figure~\ref{fig:main}{(a)-(b)}. The experiments record the evolution of the wave amplitude at a fixed position, and the time-period of the resulting signal, after transients die out, is defined to be $T_\mathrm{sat}$. The saturated amplitude $A_{\mathrm{sat}}$ is also recorded (defined as half the crest-to-trough distance as in \citet{BartheletCharruFabre1995}). The model tracks $H(0,t)$ and $H_{\mathrm{sat}}$ and calculates the corresponding $T_{\mathrm{sat}}$. The results depict the normalised values $A/A_{\mathrm{sat}}$ and $H/H_{\mathrm{sat}}$ with $T_{\mathrm{sat}}$ taken from the experiments with values $5.3$ s and $2.7$ s, approximately. The model computed values are a bit smaller (about $4.0$ s and $2$ s, respectively) as also pointed out in \cite{KalogirouCimpeanuKeavenyPapageorgiou2016R1}.The difference may be attributed to additional damping due to stable density stratification effects---see \cite{Kalogirou2018}.


The bistable travelling wave profiles shown in figure~\ref{fig:main}{(a)-(b)} are now tested for stability. We focus on initial conditions in the region between branch 1 and branch 2, rather than on relatively large or small initial data. The initial conditions are
\[
H_0(x;\theta)\;=\;(1-\theta)\,H_1\;+\;\theta\,H_2,\qquad \theta\in[0,1],
\]
where $H_1$ and $H_2$ are the branch 1 and 2 solutions. We define the basin of attraction of branches 1 and 2 to be the subinterval of $\mathcal{B}_{1,2}(\theta)\subset [0,1]$ for which the initial condition returns to the corresponding branch solution, i.e.
$\mathcal{B}_{1,2}(\theta)=\{\theta\in[0,1]:\limsup_{t\to\infty}
\|\,H(\cdot,t;H_0(\theta)) - H_{1,2}\,\|=0\}$. When both waves are stable, there exists a threshold \(\theta^{*}\in(0,1)\) such that
$\mathcal{B}_1=[0,\theta^{*}),\,
\mathcal{B}_2=(\theta^{*},1]$
and the separatrix corresponds to \(\theta=\theta^{*}\), i.e.
\(\theta^{*}=\sup\mathcal{B}_1=\inf\mathcal{B}_2\).

Figure~\ref{fig:main}(c) shows the computed basins of attraction for $\Lambda \in [0.010, 0.036]$. The value of $\theta^{*}$ is estimated to three decimal places by comparing the $L^2$ norm of the travelling wave solutions (see figure~\ref{fig:bistability}{(b)}) and the $L^2$ norm (from the dynamics) after a sufficiently large time. Branch 1 remains an attractor until $\Lambda = 0.040$. Its basin of attraction shrinks as $\Lambda$ increases, while that of branch 2 increases. However, once $\Lambda$ reaches $0.036$, the basin of branch 2 collapses to zero: branch 2 becomes unstable and a new steady unimodal branch bifurcates from it and replaces it as an attractor, which we discuss in \S \ref{sec:newbranch}.


\subsection{New symmetry-breaking travelling waves and further bistability}\label{sec:newbranch}

\begin{figure}
  \centering

  \begin{minipage}{0.40\linewidth}
    \begin{overpic}[width=\linewidth]{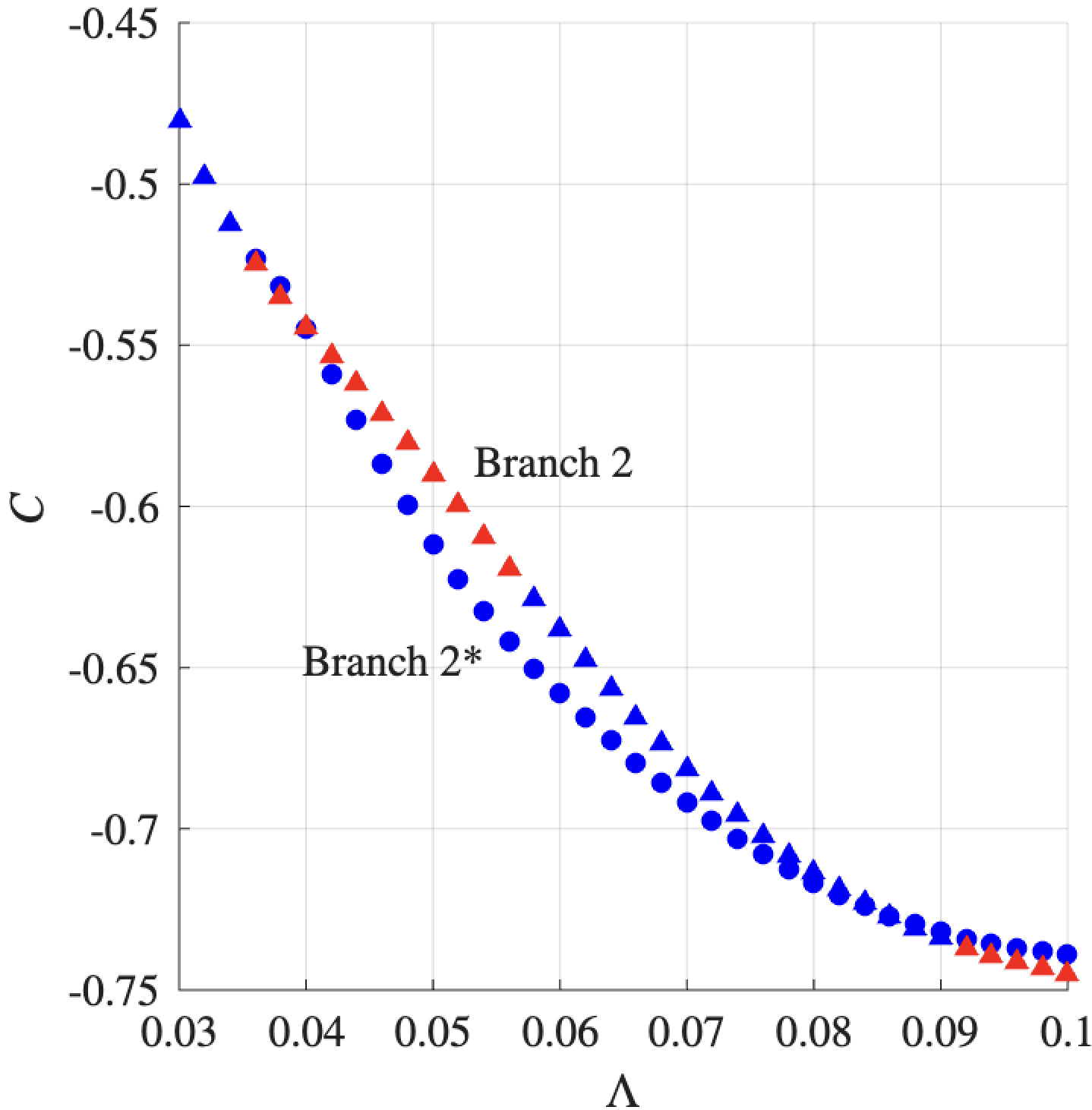} 
      \put(0,99){(\textit{a})}
    \end{overpic}
  \end{minipage}\hspace{0.02\textwidth}
  \begin{minipage}{0.40\linewidth}
    \begin{overpic}[width=\linewidth]{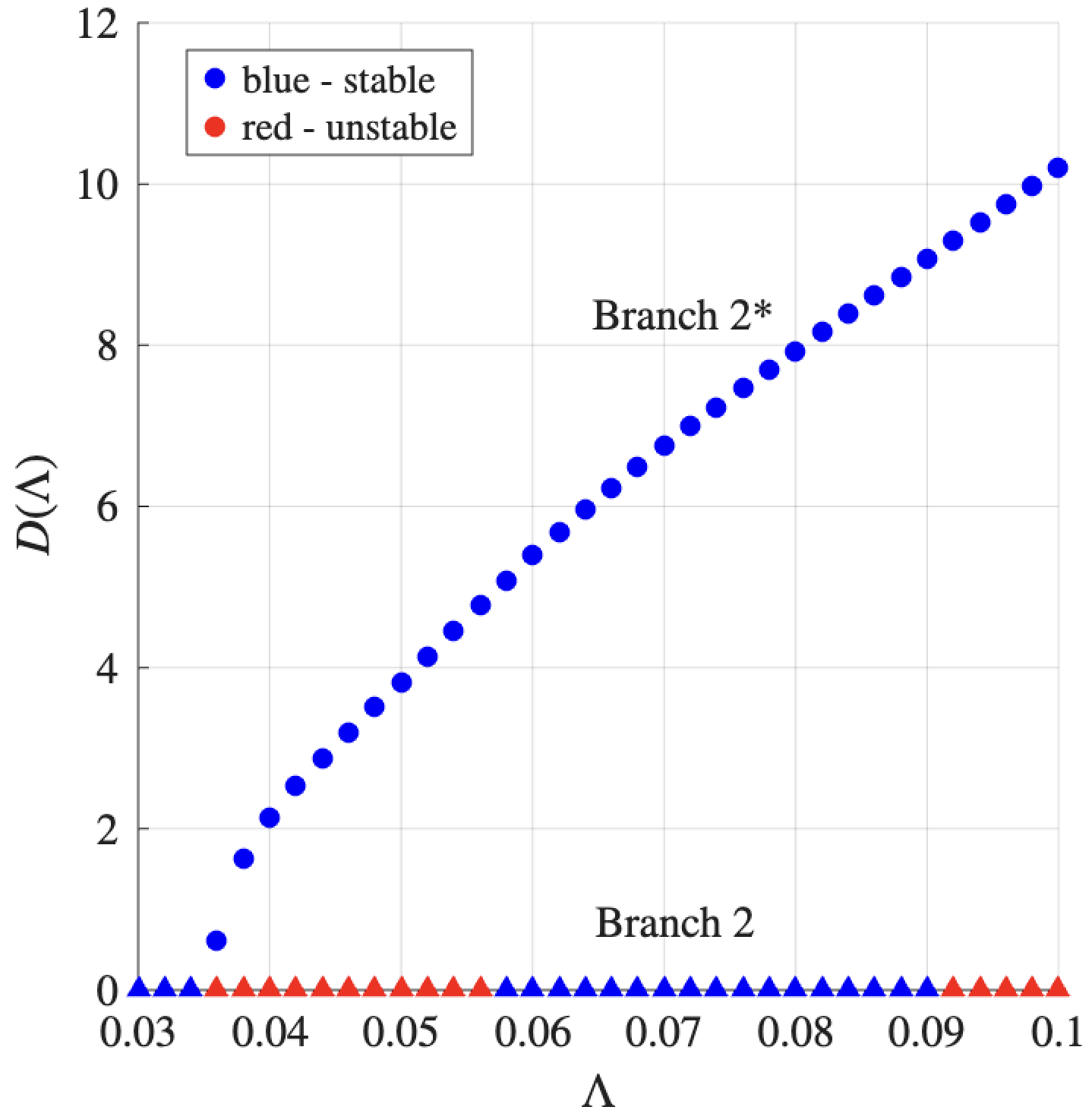} 
      \put(0,99){(\textit{b})}
    \end{overpic}
  \end{minipage}
  
  \begin{minipage}{0.40\linewidth}
    \begin{overpic}[width=\linewidth]{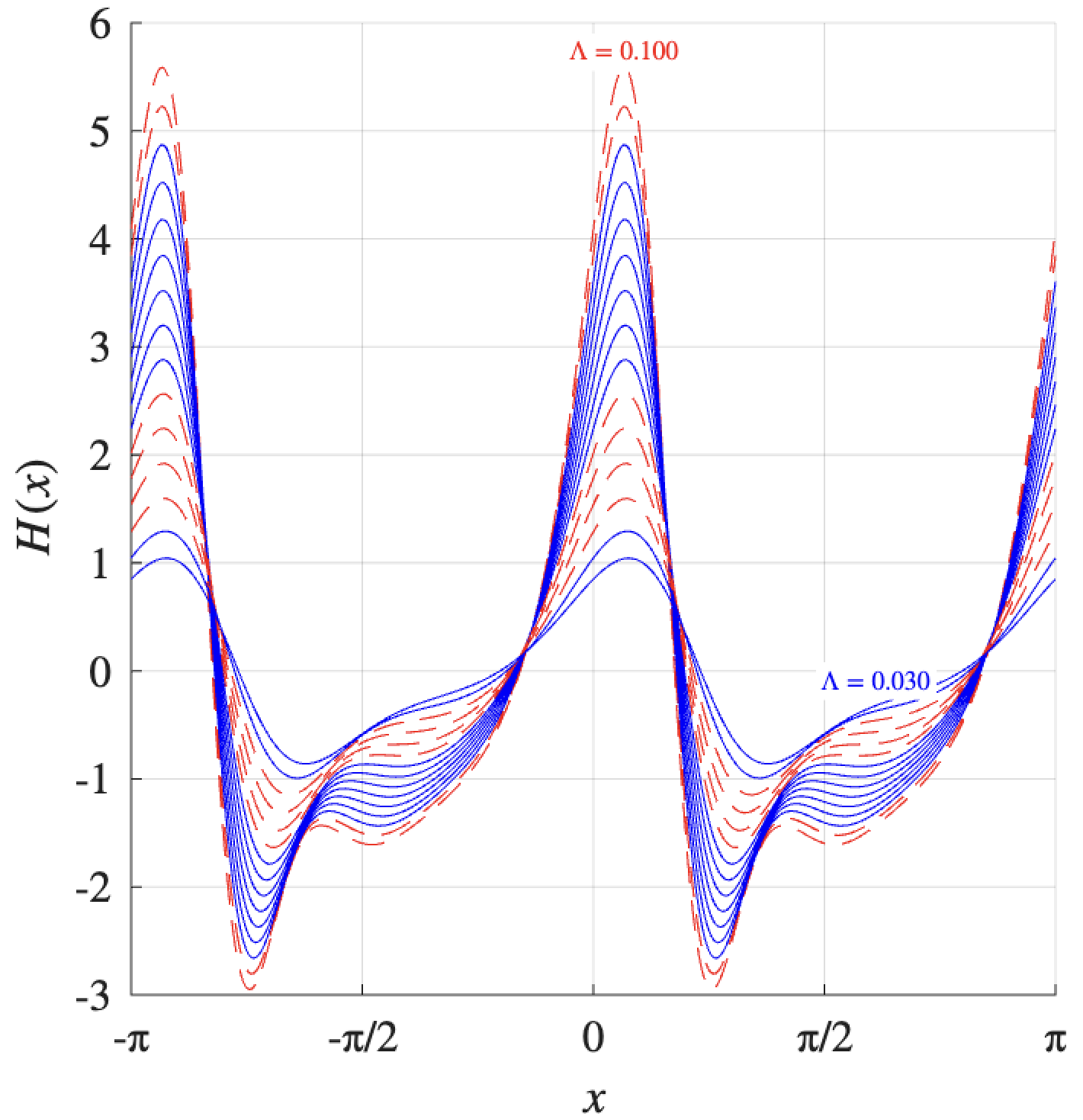} 
      \put(0,99){(\textit{c})}
    \end{overpic}
  \end{minipage}\hspace{0.02\textwidth}
  \begin{minipage}{0.40\linewidth}
    \begin{overpic}[width=\linewidth]{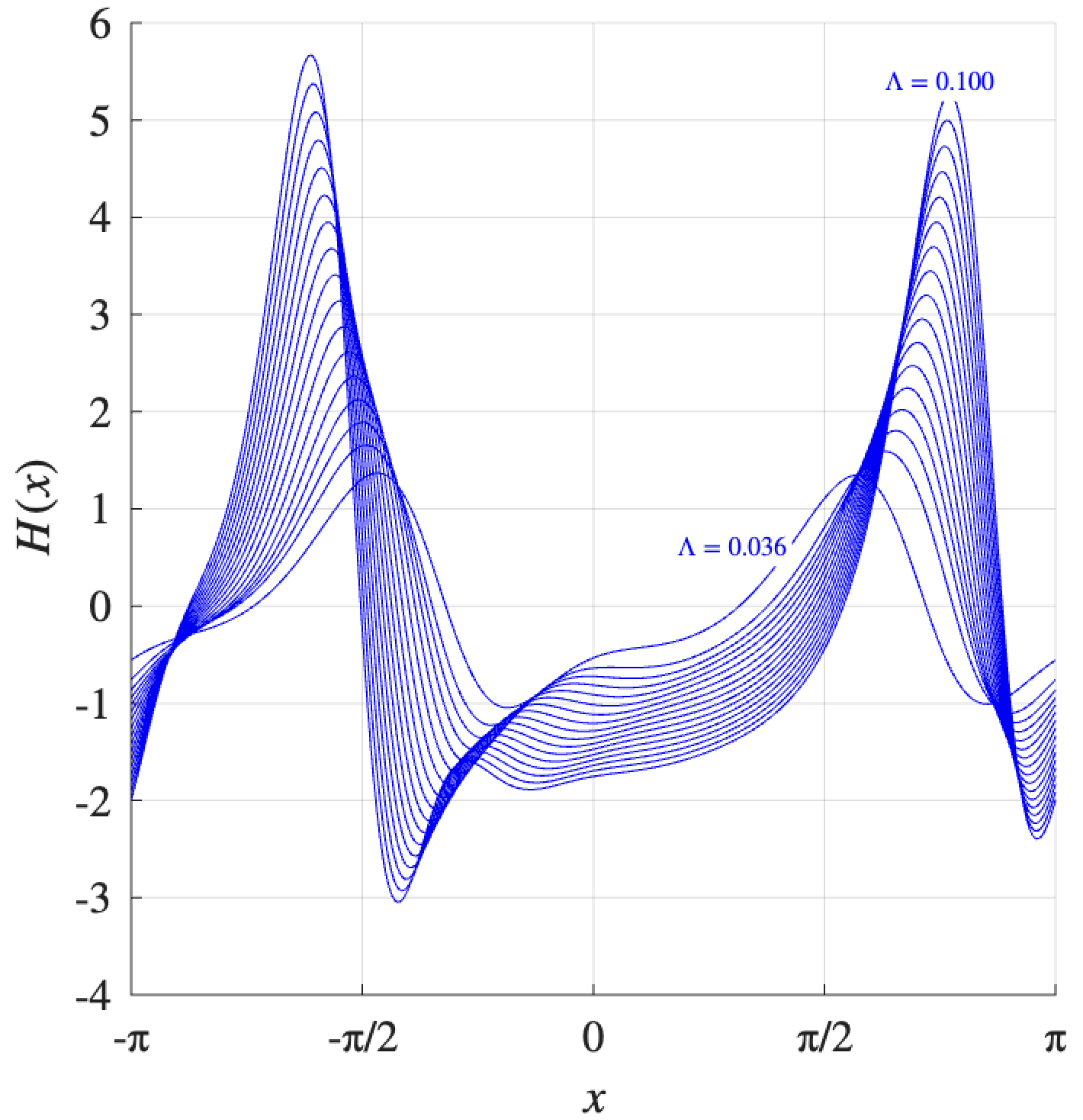} 
      \put(0,99){(\textit{d})}
    \end{overpic}
  \end{minipage}

  \begin{minipage}{0.40\linewidth}
    \begin{overpic}[width=\linewidth]{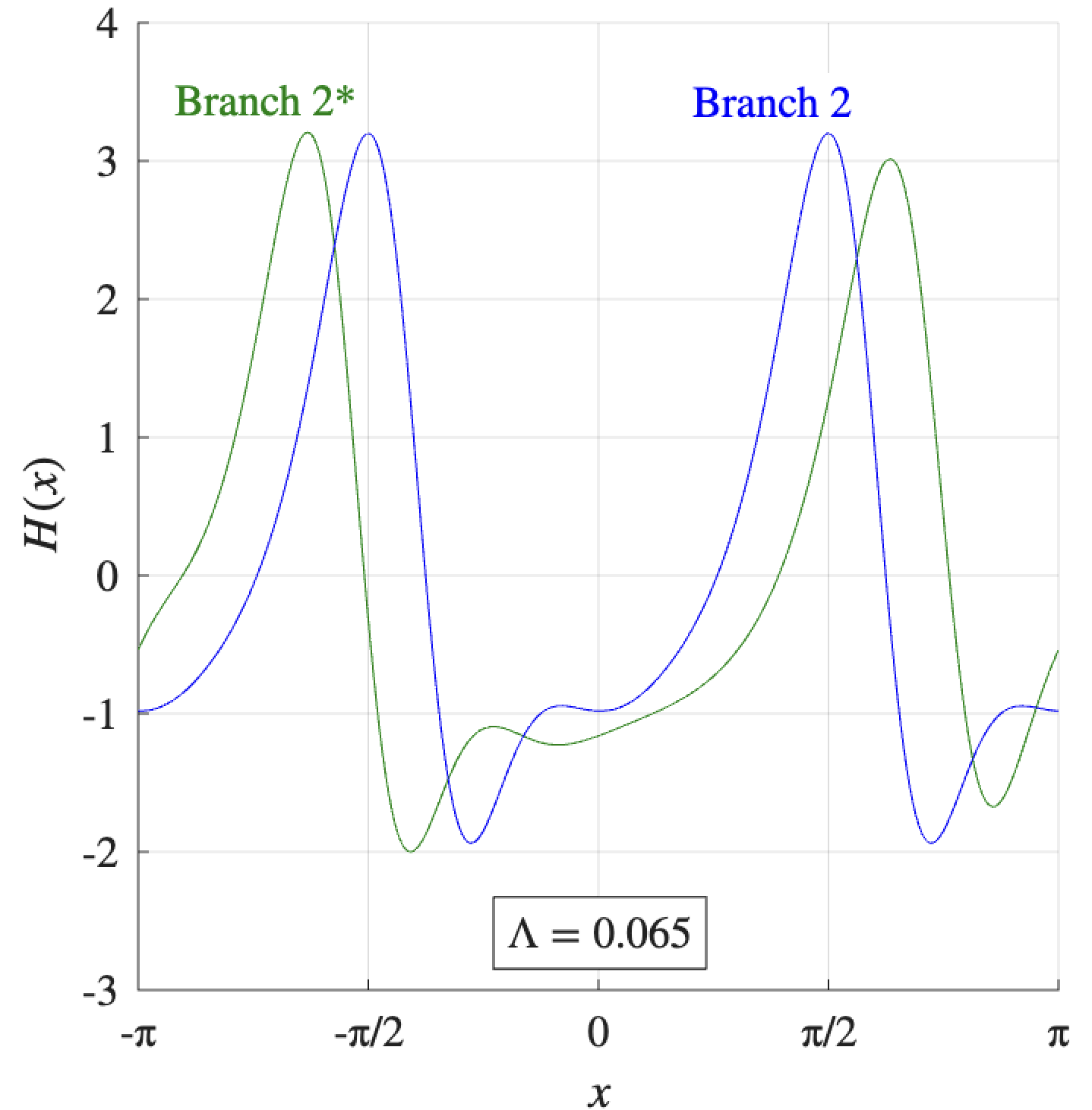} 
      \put(0,99){(\textit{e})}
    \end{overpic}
  \end{minipage}\hspace{0.02\textwidth}
  \begin{minipage}{0.40\linewidth}
    \begin{overpic}[width=\linewidth]{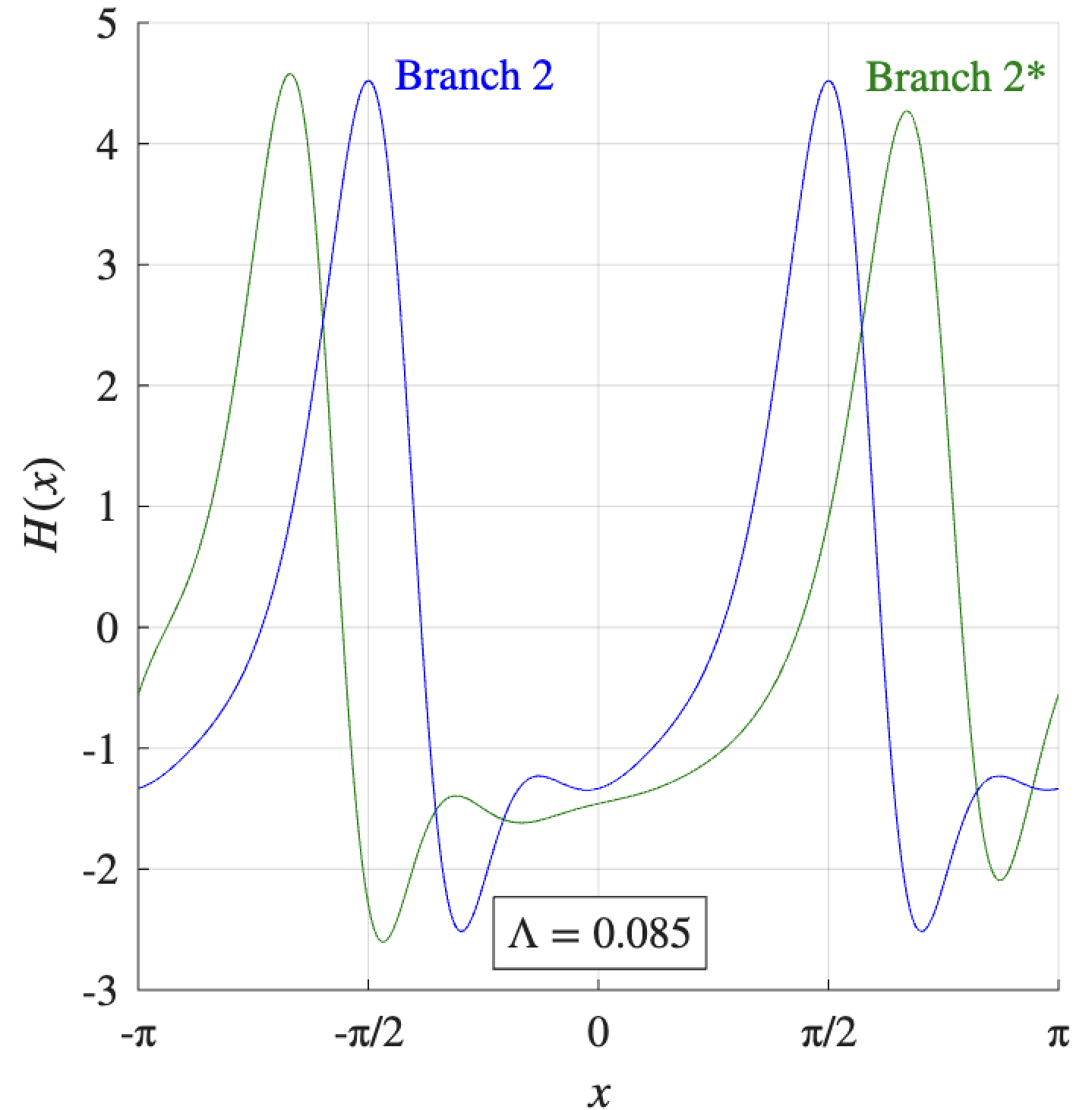} 
      \put(0,99){(\textit{f})}
    \end{overpic}
  \end{minipage}
  \caption{Bifurcation diagrams of branch~2 and branch~2$^\ast$.\quad
(\textit{a}) Wave speed $C$ versus $\Lambda$;\quad
(\textit{b}) half-period shift norm $D(\Lambda)$ versus $\Lambda$;\quad
(\textit{c}) wave profiles for branch~2 (bimodal);\quad
(\textit{d}) wave profiles for branch~2$^\ast$ (symmetry-broken);\quad
(\textit{e})-(\textit{f}) comparison of travelling wave profiles for branch 2 (blue, labelled) and branch~2$^\ast$ (green, labelled)
for $\Lambda=0.065$ and 
$\Lambda=0.085$; the solutions are shifted to be crest-symmetrised about $x=0$.}
  \label{fig:newbranch}
\end{figure}

We now focus on this new branch, which has not been observed either in \citet{BartheletCharruFabre1995} or in \citet{KalogirouCimpeanuKeavenyPapageorgiou2016R1}. However, this new branch is important as it stays as an attractor locally or globally for a large range of $\Lambda$. Figure~\ref{fig:newbranch}(a) displays the bifurcation diagram in terms of the phase speed $C$ versus $\Lambda$. Representative profiles for branch~2 and for the symmetry-broken branch are shown in figures~\ref{fig:newbranch}(c) and~\ref{fig:newbranch}(d), respectively. Since the $L^2$-norm of branch 2 and the new branch are too close to each other, a better diagnostic is introduced in figure~\ref{fig:newbranch}(b) that uses the half-period shift norm
\begin{equation}
D(\Lambda)\;=\;\left(\int_{-\pi}^{\pi}\bigl|H(z;\Lambda)-H(z+\pi;\Lambda)\bigr|^{2}\,dz\right)^{1/2}.\label{eq:DLambda}
\end{equation}
This value quantifies departure from $\pi$-periodicity. In particular, $D(\Lambda)=0$ for perfectly $\pi$-periodic bimodal waves (branch~2), whereas $D(\Lambda)>0$ indicates symmetry breaking and an effectively $2\pi$-periodic waveform, which we denote as branch~2$^\ast$.

At $\Lambda \approx 0.036$, branch~2 undergoes a steady bifurcation: the dominant eigenvalue of the linearised operator about the travelling wave crosses the imaginary axis at zero frequency. Past this point branch~2 becomes unstable, and branch~2$^\ast$ emerges that breaks the $\pi$-period symmetry. Figure~\ref{fig:newbranch}(d)(e)(f) show that this new branch is no longer $\pi$-periodic: the two peaks of the bimodal state become unequal. Meanwhile, figure~\ref{fig:newbranch}(b) tracks the half-period shift norm $D(\Lambda)$. Branch~2$^\ast$ has $D(\Lambda)>0$ and its defect grows monotonically with $\Lambda$, reflecting the increasing inequivalence of the two peaks. This symmetry-breaking, together with the steady (real-eigenvalue) crossing, is consistent with a pitchfork bifurcation at $\Lambda\approx 0.036$ for branch~2. 

Because the governing  equation \ref{eq:tw-ode} is translation invariant, if $H(z)$ is a travelling-wave solution then so is $H(z+z_0)$ for any shift $z_0$. In particular, the symmetry-breaking bifurcation from the $\pi$-periodic branch~2 gives rise to a pair of symmetry-related $2\pi$-periodic solutions that are mapped into each other by the half-period shift $z\mapsto z+\pi$. In the present computations for branch~2 we fix the phase by imposing $\Im\,\widehat H_{1}=0$, which selects a single representative from this shift orbit. We have also verified numerically the existence of the symmetry-related partner. In the representative branch~2$^\ast$ profiles shown in figure~\ref{fig:newbranch}(d), the two crests become increasingly unequal and drift away from $x=0$ as $\Lambda$ increases, whereas in the $\pi$-shifted partner the corresponding crests drift towards $x=0$. The two members of the pair are physically equivalent and coincide in scalar diagnostics such as $C$, $\|H\|_{L^2}$ and the defect measure $D(\Lambda)$; the difference is only in their phase. 

As $\Lambda$ is increased further, branch~2 experiences a second steady bifurcation at $\Lambda\approx 0.057$. Within the interval $\Lambda\in[0.057,0.090]$ both branch~2 and branch~2$^\ast$ are stable, yielding bistability. At $\Lambda\approx 0.090$, a third steady bifurcation takes place and branch~2 turns unstable again. Branch~2$^\ast$ remains stable until $\Lambda \approx 0.137$, where it loses stability through a Hopf bifurcation.

\subsection{Global structure and periodic orbits}\label{sec:overview}

\begin{figure}
  \centering

  \begin{minipage}{0.74\linewidth}
    \begin{overpic}[width=\linewidth]{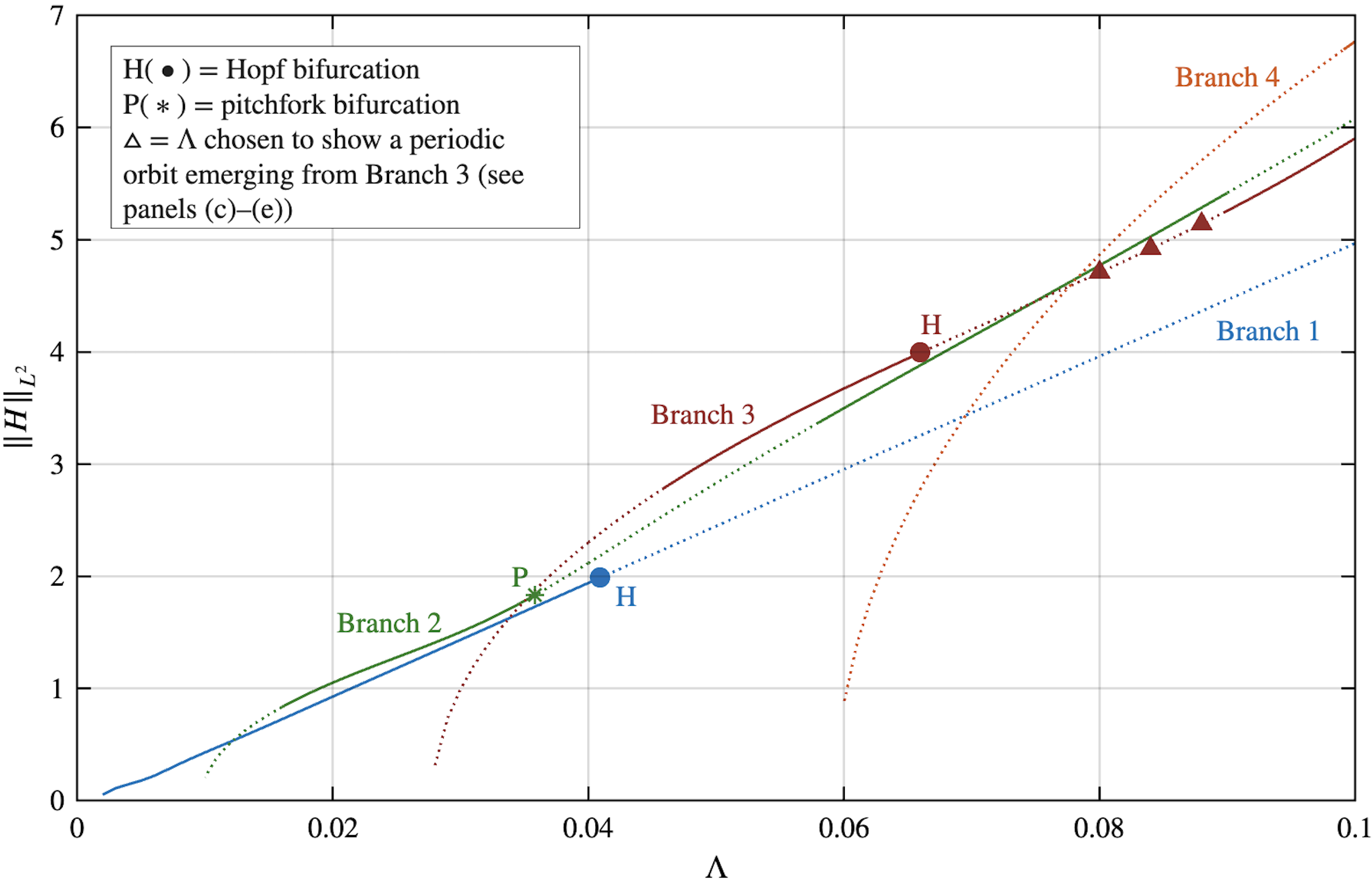}
      \put(0,64){(\textit{a})}
    \end{overpic}
  \end{minipage}

  \vspace{0.8em}

  \begin{minipage}{0.37\linewidth}
    \begin{overpic}[width=\linewidth]{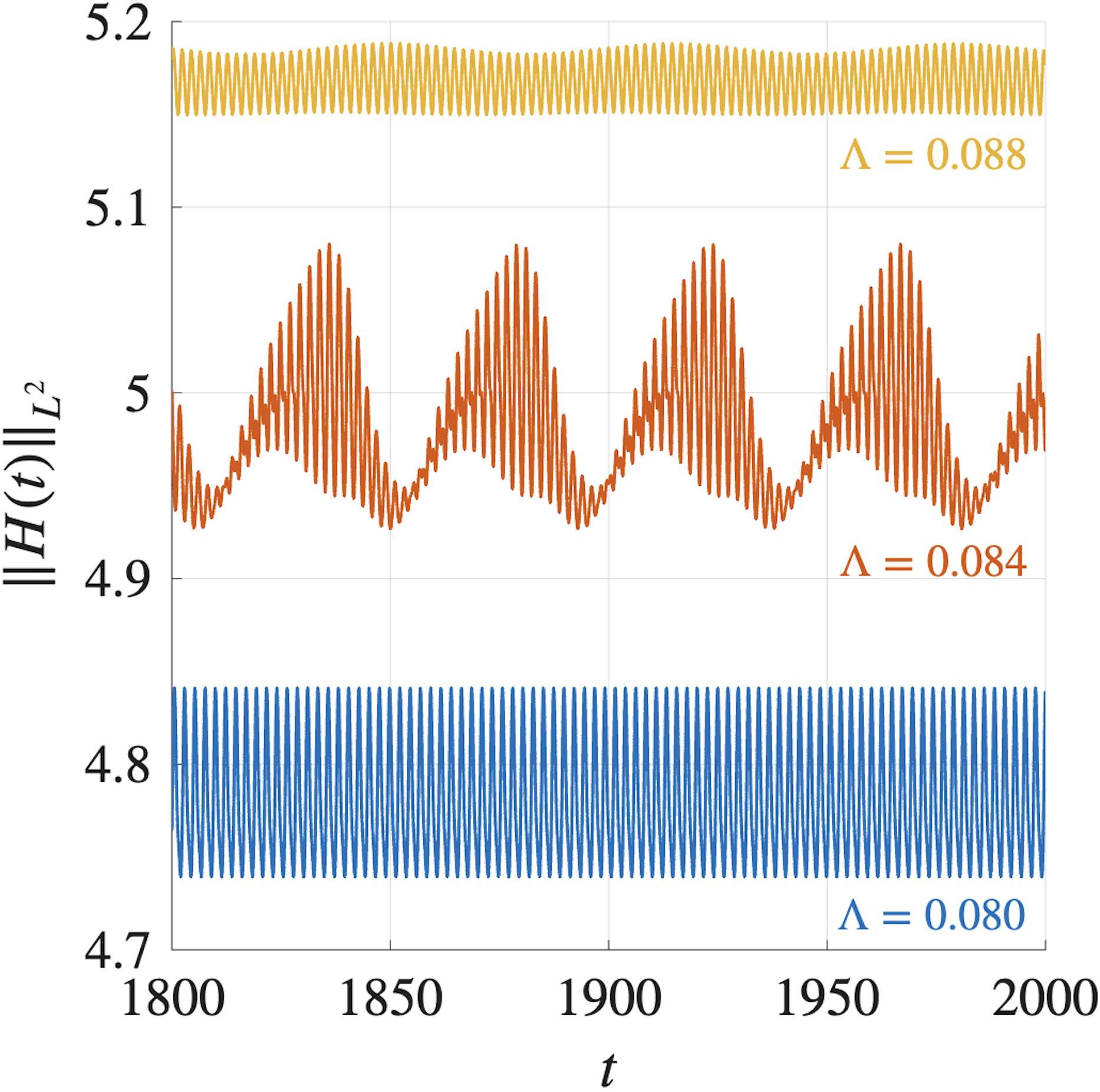} 
      \put(0,99){(\textit{b})}
    \end{overpic}
  \end{minipage}\hspace{0.02\textwidth}
  \begin{minipage}{0.37\linewidth}
    \begin{overpic}[width=\linewidth]{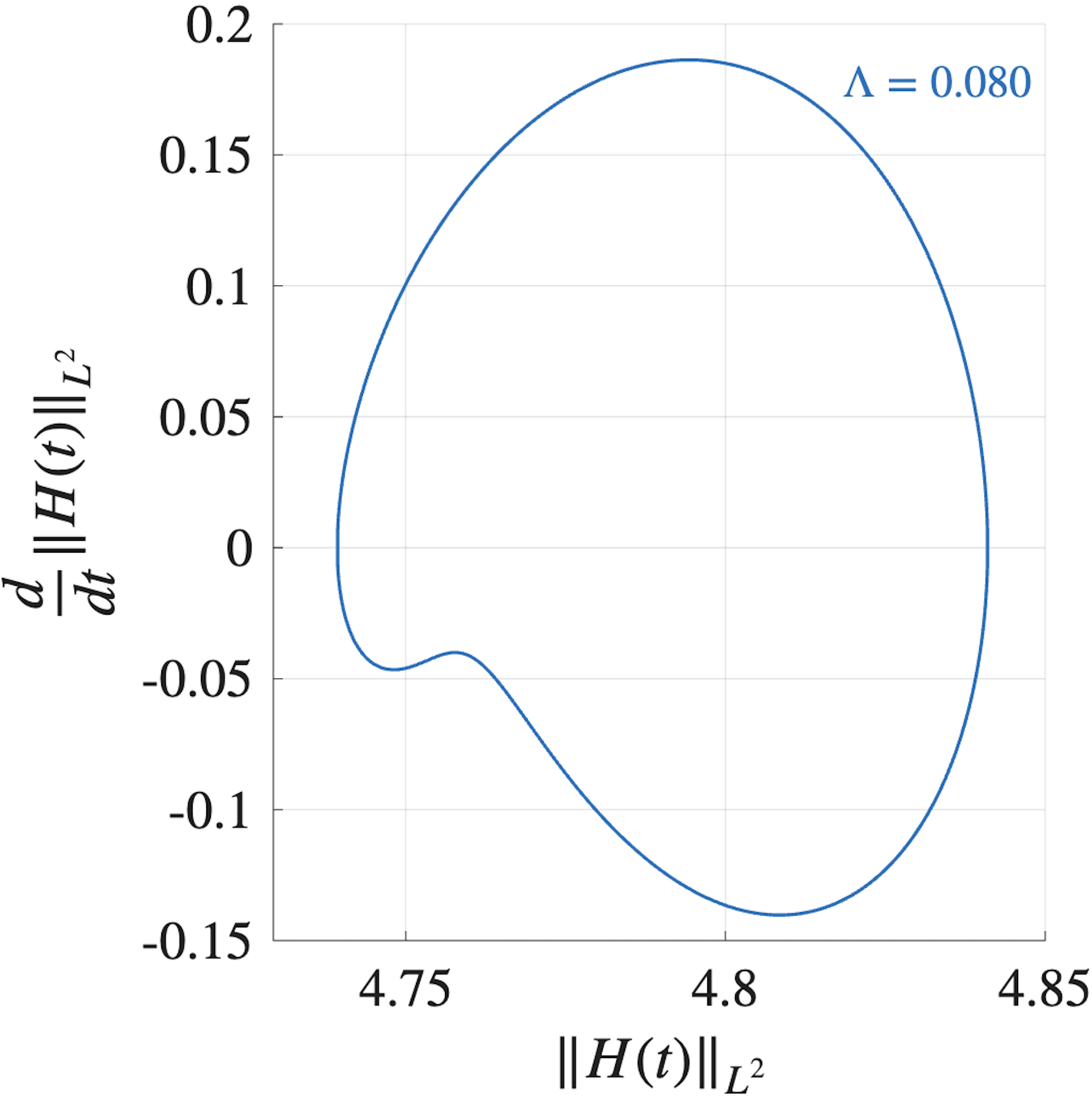} 
      \put(0,99){(\textit{c})}
    \end{overpic}
  \end{minipage}

  \begin{minipage}{0.37\linewidth}
    \begin{overpic}[width=\linewidth]{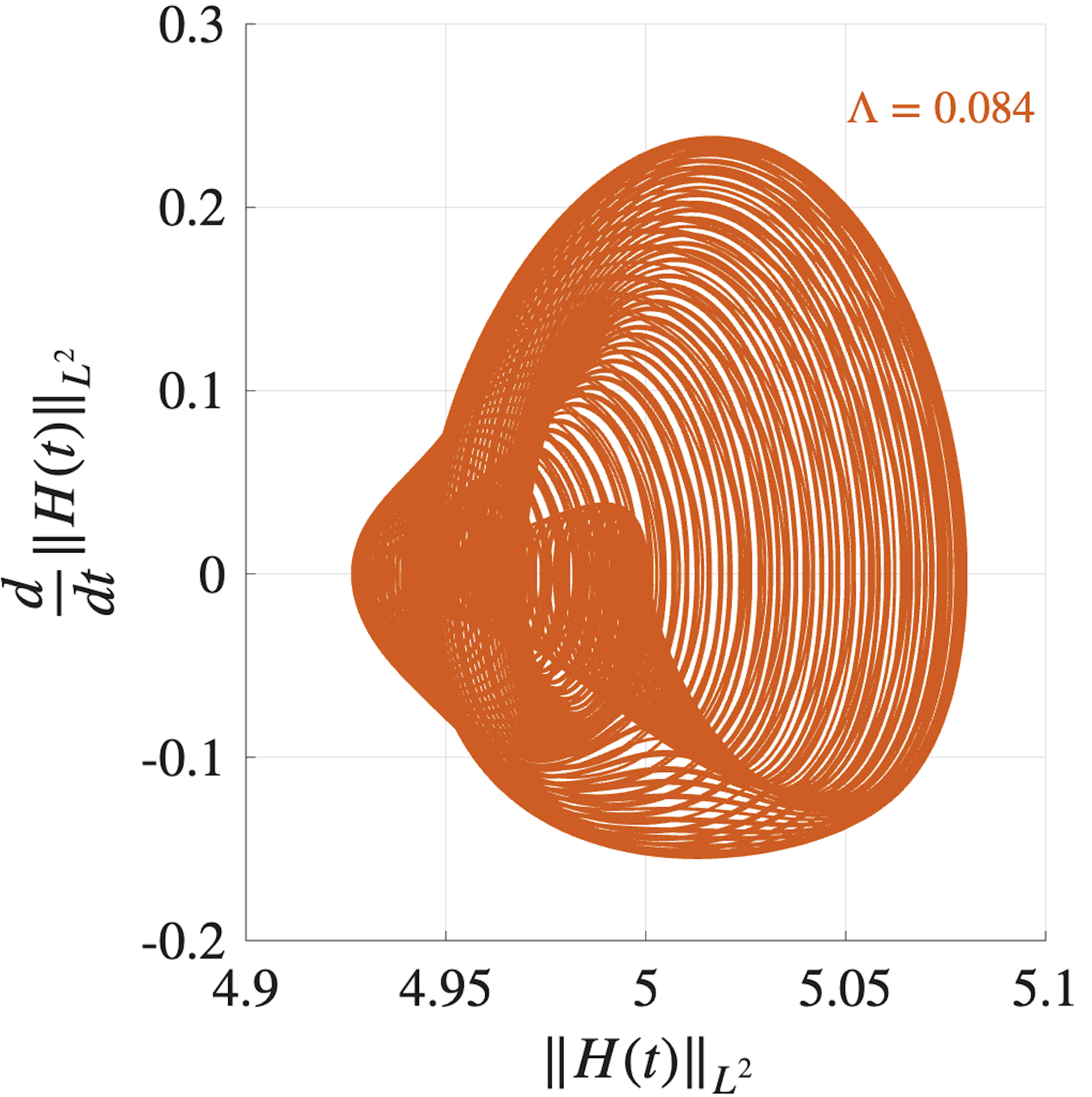} 
      \put(0,99){(\textit{d})}
    \end{overpic}
  \end{minipage}\hspace{0.02\textwidth}
  \begin{minipage}{0.38\linewidth}
    \begin{overpic}[width=\linewidth]{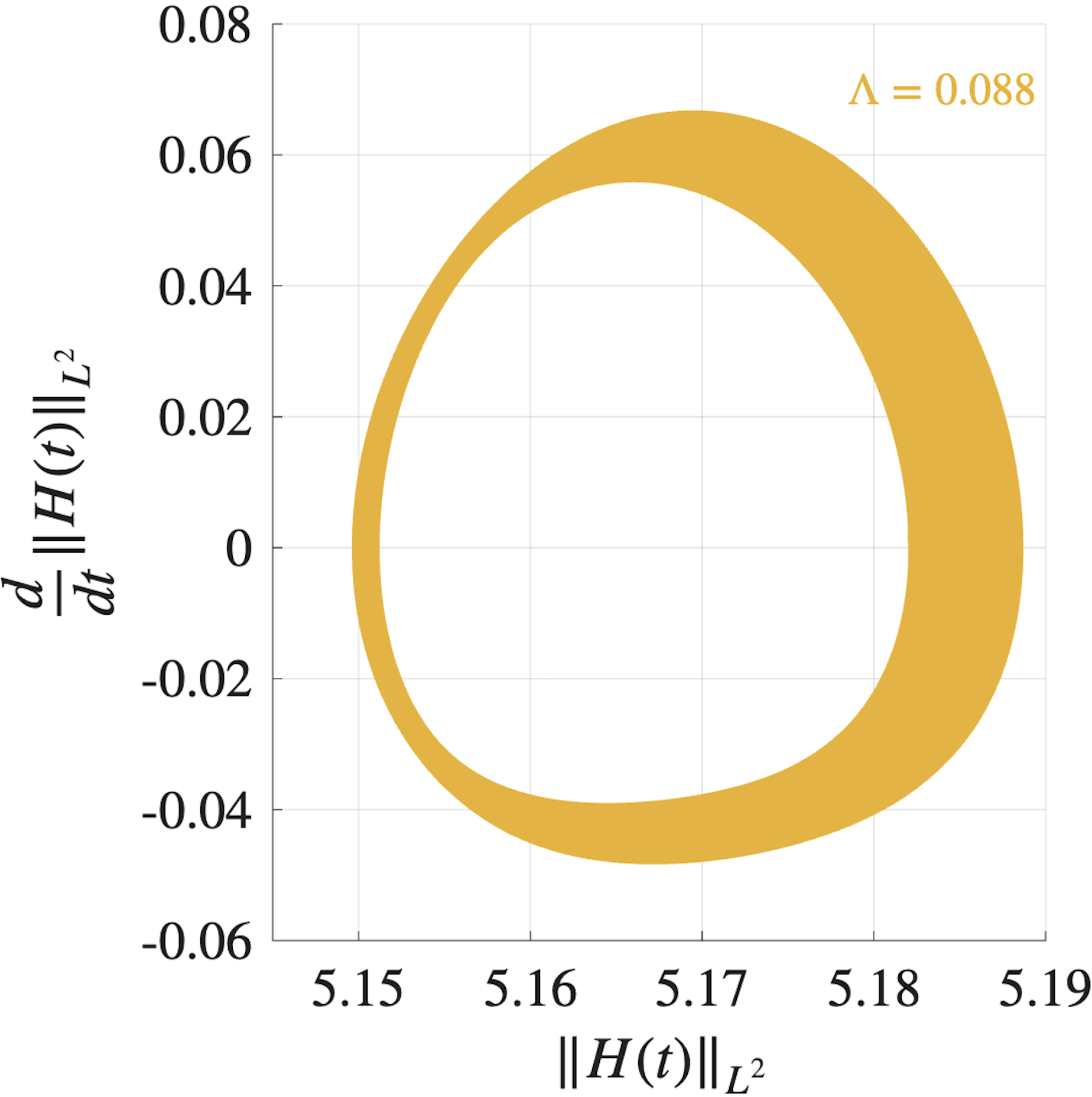} 
      \put(0,99){(\textit{e})}
    \end{overpic}
  \end{minipage}

  \caption{(\textit{a}) Global bifurcation diagram ($L^2$-norm versus $\Lambda$) for branches 1--4 (except branch 2*) with solid segments stable and dotted segments unstable; \quad (\textit{b}) dynamics of $L^2$-norm after saturation for $\Lambda = 0.080, 0.084$ and $0.088$ (also labelled in (\textit{a}));\quad (\textit{c})--(\textit{e}) corresponding phase portrait in the plane $(\|H(t)\|_{L^2},\,\frac{d}{dt}\|H(t)\|_{L^2})$.}
  \label{fig:overview}
\end{figure}

We now map other higher-wavenumber travelling-wave families and show a few examples of time-dependent attractors for $\Lambda \in [0, 0.100]$. Figure~\ref{fig:overview} summarises the bifurcation diagram in the $(\Lambda,\|H\|_{L^{2}})$ plane for several families of spatially periodic travelling waves of \eqref{eq:tw-ode}. Here branch $k$ denotes a $2\pi/k$-periodic travelling wave that bifurcates from a flat state as a linear mode with wavenumber $k$ (equivalently, it is supported on Fourier harmonics that are integer multiples of $k$). Translation invariance is removed by fixing the phase through $\Im\,\widehat H_{k}=0$ as discussed earlier. In addition to the fundamental $2\pi$-periodic family (branch~1) and the $\pi$-periodic bimodal family (branch~2), the system supports multimodal travelling waves, namely branch~3 ($2\pi/3$-periodic) and branch~4 ($\pi/2$-periodic), whose $\|H\|_{L^{2}}$ typically exceeds that of branches~1 and 2 at the same $\Lambda$. The symmetry-broken branch~2$^\ast$ is not included in the plot since its $\|H\|_{L^{2}}$ norm is nearly identical to that of branch~2 over the range of interest---the start of branch $2^*$ is identified by a star symbol in figure \ref{fig:overview}(a) at the pitchfork bifurcation point $P$. Stability changes of branches~3 and 4 occur exclusively through Hopf bifurcations: branch~3 possesses two disjoint stability windows, becoming stable near $\Lambda\approx 0.046$, losing stability near $\Lambda\approx 0.066$, restabilising near $\Lambda\approx 0.090$, and destabilising again near $\Lambda\approx 0.104$. Intriguingly, therefore, our computations reveal the possibility of tristability where stable solutions from branches 2, 2$^\ast$ and 3 can co-exist for $0.057\lesssim \Lambda \lesssim 0.066$. On the other hand, branch~4 is stable on a single interval $0.099\lesssim \Lambda \lesssim 0.135$. Consequently, for $\Lambda\gtrsim 0.060$ the model admits multiple coexisting travelling-wave families of increasing spatial complexity, whereas beyond the Hopf bifurcation of branch~2$^\ast$ at $\Lambda\approx 0.137$, no steady travelling waves among the branches shown remain linearly stable.

To connect the branch structure in figure~\ref{fig:overview}(a) with the evolution PDE, we also monitor long-time dynamics in the parameter range where branch~3 is linearly unstable (after its Hopf bifurcation at $\Lambda\approx 0.066$ and before $\Lambda\approx 0.090$, where it turns stable). All computations in this time-periodic regime are initiated from the small-amplitude mixed-mode perturbation
\[
H(x,0)=10^{-3}\sin(3x)+10^{-3}\sin(4x),
\]
chosen to excite the $k=3$ dominated branch while allowing interaction with its adjacent harmonics. Figure~\ref{fig:overview}(b) shows saturated time traces of $\|H(t)\|_{L^{2}}$ for $\Lambda=0.080,0.084$ and $0.088$ (marked by filled triangles in figure~\ref{fig:overview}(a)), illustrating a non-monotone evolution of the post-transient dynamics: modulation strengthens as $\Lambda$ increases away from the Hopf point, but collapses again as $\Lambda$ approaches the restabilisation of branch~3 near $\Lambda\approx 0.090$. To visualise the post-transient dynamics in a low-dimensional projection, we plot the phase plane $(\|H(t)\|_{L^2},\,\frac{d}{dt}\|H(t)\|_{L^2})$ in figure~\ref{fig:overview}(c)-(e). The quantity $\frac{d}{dt}\|H(t)\|_{L^2}$ is computed with spectral accuracy by multiplying \eqref{eq:main} by $H(x,t)$, integrating over a period and using Parseval's theorem to calculate integrals in Fourier space.
At $\Lambda=0.080$ the solution is time-periodic with a single fundamental frequency as confirmed by the phase plane in figure~\ref{fig:overview}(c). Increasing $\Lambda$ to $0.084$, introduces long and short modulations to the $\|H(t)\|_{L^{2}}$ signal as seen in figure \ref{fig:overview}(b); there is strong numerical evidence from the phase-plane \ref{fig:overview}(d) that the dynamics is quasi-periodic---return maps can establish this as in similar studies by \cite{Akrivis-et-al-IMAJNA}, \cite{Kalogirou-et-al-2012}, \cite{KalogirouPapageorgiou2016}. Increasing $\Lambda$ further to $\Lambda=0.088$, and hence closer to the re-stabilisation bifurcation point, the oscillations remain modulated with a weaker envelope as seen in the evolution of $\|H(t)\|_{L^2}$ in figure \ref{fig:overview}(b); the behaviour again appears to be quasi-periodic as evidenced by the phase plane in figure \ref{fig:overview}(e). As $\Lambda$ increases to values close to $\Lambda=0.090$, the loops in the phase-plane are expected to decrease in size and produce a fixed point which is the genesis of the stable travelling waves on branch 3 for $\Lambda>0.090$.


The time-dependent computations also reveal other interesting phenomena, including metastability in multistable regimes (figure~\ref{fig:overview}(a)) and time-periodic orbits at larger $\Lambda$.  We observe metastability associated with both branch~2 and branch~4: depending on the initial condition, trajectories can approach these travelling-wave levels rapidly and then drift away only slowly along weakly unstable directions. In addition, Hopf destabilisations of branches 3 and 4 are accompanied by time-periodic solutions dominated by the corresponding spatial mode: for branch~3, such $k=3$ dominated time-periodic orbits occur in the interval where branch~3 is unstable via a Hopf bifurcation at $\Lambda\approx 0.066$, and analogous time-periodic solutions are observed in simulations again when branch~3 loses stability near $\Lambda\approx 0.104$ (Hopf) and when branch~4 loses stability near $\Lambda\approx 0.135$ (Hopf), with dynamics dominated by modes $k=3$ and $k=4$, respectively. A detailed continuation and stability analysis of these higher-$\Lambda$ time-periodic solutions is left for future work.

Overall, within $\Lambda\in[0,0.100]$ we identify the travelling-wave family branches~1--4, the symmetry-broken branch~2$^{*}$, and time-periodic orbits emerging from branch~3. To our knowledge, the higher-wavenumber families (branches~3--4), the symmetry-breaking branch~2$^{*}$, and the accompanying periodic dynamics have not been reported previously for this model. These results provide concrete guidance for further experiments in the same physical setting, where multiple distinct coherent states and time-dependent attractors can coexist and can be selected by carefully choosing the initial disturbance.

\section{Extensive comparison with the experiments \cite{BartheletCharruFabre1995}}\label{sec:fullcomparison}

In this section, we present a full comparison between the experiment and the model of all the cases having $d = 0.25$, except for the bistability phenomena already addressed in \S ~\ref{sec:comparison}.

\subsection{Interfacial profiles}
\label{sec:IP}

\begin{figure}
  \centering
  \begin{minipage}{0.40\textwidth}
    \begin{overpic}[width=\linewidth]{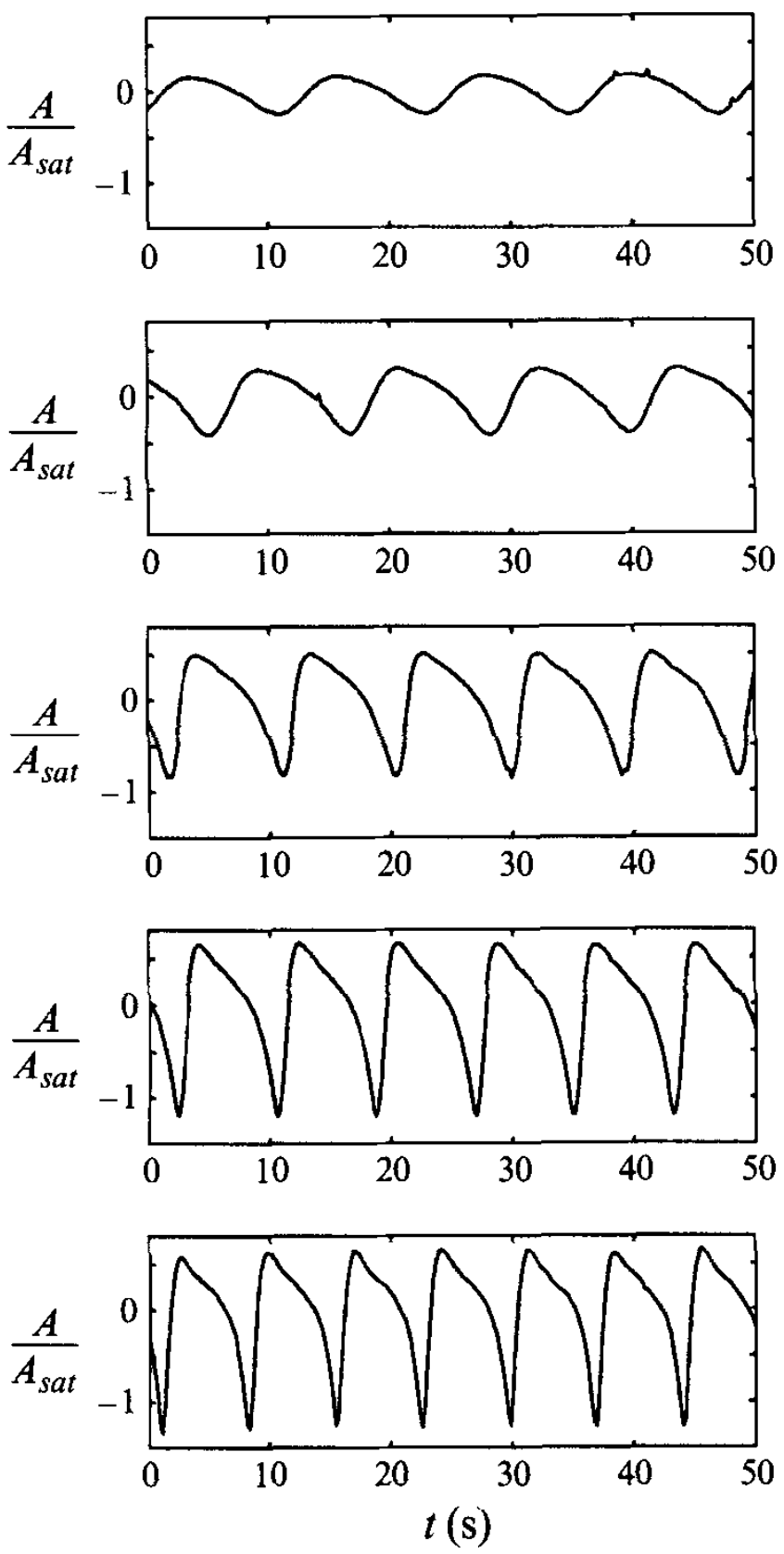}
      \put(0,99){(\textit{a})}
    \end{overpic}
\end{minipage}    \hspace{0.02\textwidth}
  \begin{minipage}{0.40\textwidth}
    \begin{overpic}[width=\linewidth]{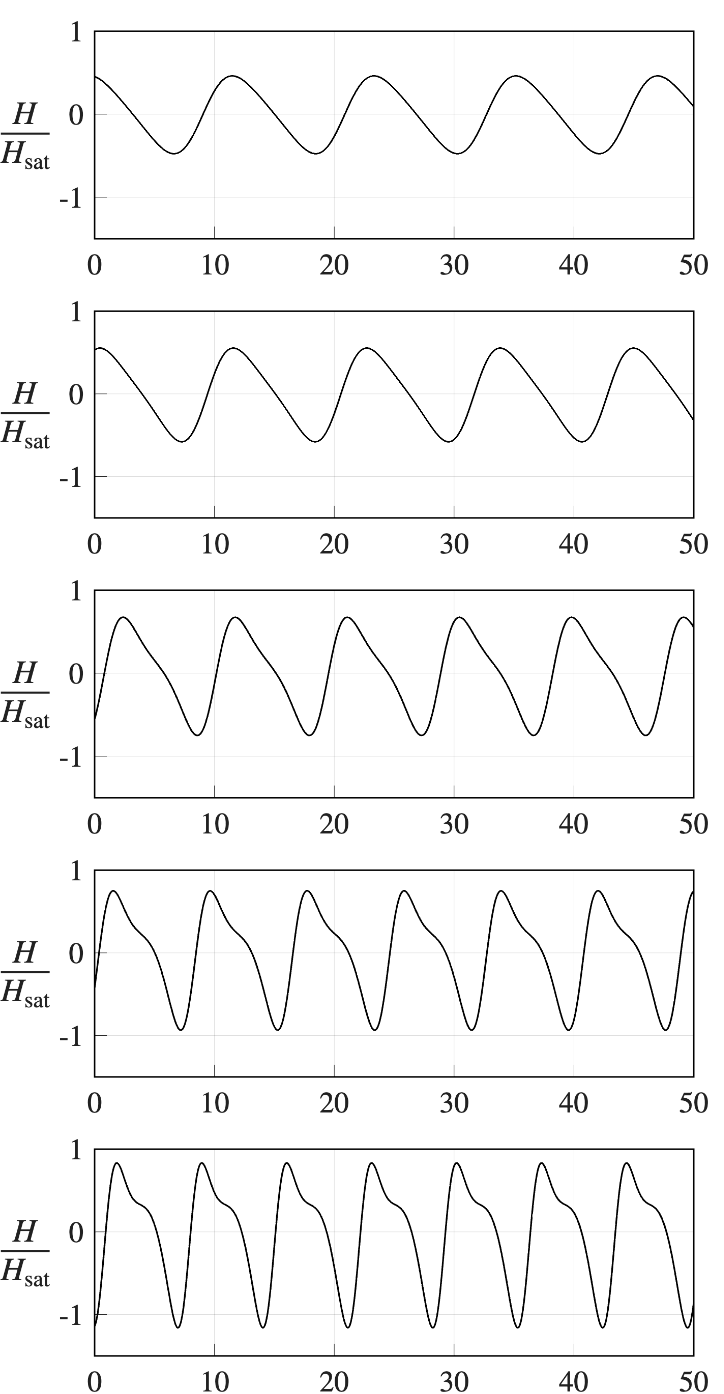}
      \put(0,98){(\textit{b})}
    \end{overpic}
    \vspace{0.3em}\makebox[\linewidth]{\fontsize{10}{12}\selectfont $t$ (s)}

  \end{minipage}

  \caption{Interfacial profile for $U/U_L= \{1.06, 1.13,1.33,1.58, 1.77\}$. \( U_L = 0.138~\mathrm{m\,s^{-1}} \).\quad (a) Experimental shapes (from \citet{BartheletCharruFabre1995}, p.36);\quad (b) Model computation. Amplitudes are normalised with the saturated amplitude for $U/U_L=1.77$.}
  \label{fig:IP}
\end{figure}

We proceed with comparisons of numerical solutions of \eqref{eq:main} with the experiments detailed in table \ref{tab:exp-phys}. For given parameters, the experiments record the plate velocity $U_L$ above which the flow becomes unstable to long waves with wavelength equal to the device perimeter. Different experiments then provide the ratio $U/U_L$, from which we calculate the Reynolds numbers $R$ to use in the model computations. All six cases were studied in detail by \citet{BartheletCharruFabre1995}. The interfacial shapes for $U/U_L = 1.33, 1.58, 1.77$, were shown in figure~3 of \citet{KalogirouCimpeanuKeavenyPapageorgiou2016R1}, but the evolution of harmonic amplitudes was not examined there. Experiments are reported for $U/U_L= 1.06, 1.13,1.33,1.58, 1.77, 2.25$, with $U_L = 0.138~\mathrm{m\,s^{-1}}$, which correspond to $R = 309.36, 329.78, 388.15$, $461.11, 516.56, 656.65$. The corresponding values of $\Lambda$ used in the model are  $0.00449, 0.00479, 0.00564, 0.00669$, $ 0.00750, 0.00953$. We fix $\Lambda=0.00750$ at $U/U_L=1.77$, where the model best matches the experimental interfacial shapes ($\Lambda$ can be thought of as a fitting parameter in matching one of the experiments---once this is done the values of $\Lambda$ for the other four experiments follow as explained next). Since $\Lambda$, $\mathrm{Ca}$ and $U$ are proportional to each other for the same fluid, we scale $\Lambda$ linearly with $U/U_L$ to obtain the other values of $\Lambda$. Within the range explored here, variations in $\Lambda$ produce only minor quantitative changes; the qualitative behaviour is governed primarily by $R$.

The agreement with the experiments is excellent. The waves in both panels of figure~\ref{fig:IP} display similar characteristics, with steep fronts and pronounced troughs that become more prominent as $R$ increases, while the oscillation period shortens with increasing $R$. The results depict the normalised values $A/A_{\mathrm{sat}}$ and $H/H_{\mathrm{sat}}$ with $T_{\mathrm{sat}}$ taken from the experiments with values $11.8$s, $11.1$s, $9.4$s, $8.1$s, $7.1$s, $6.5$s, approximately. Similar to \S \ref{sec:comparison}, the model computed values are a bit smaller. The bistability case can be seen as a continuation of the cases studied in this section, with a larger $U / U_L$. 

\subsection{Harmonic amplitudes}\label{sec:HA}

\begin{figure}
  \centering
  \begin{minipage}{0.40\textwidth}
    \begin{overpic}[width=\linewidth]{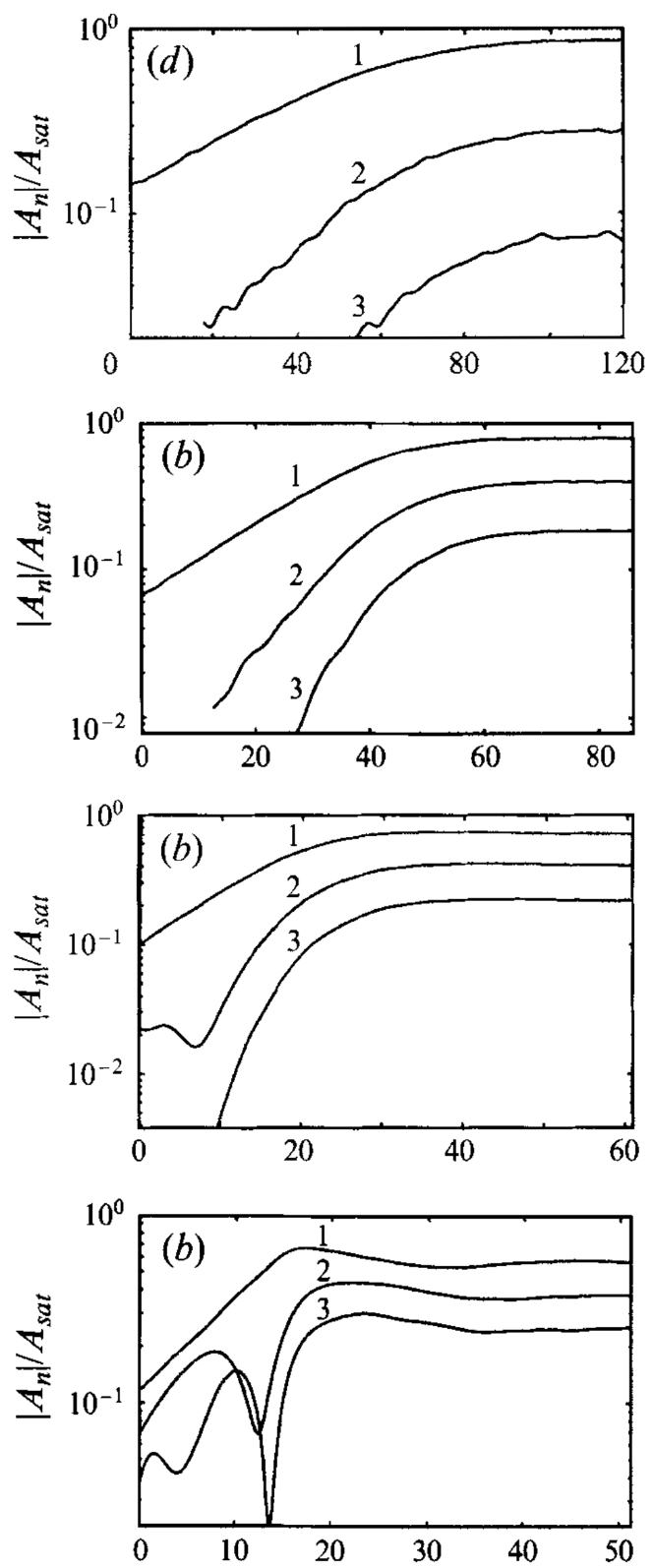}
      \put(0,98){(\textit{a})}
    \end{overpic}
    \vspace{0.3em}\makebox[\linewidth]{\fontsize{10}{12}\selectfont $t/T_{\mathrm{sat}}$}
  \end{minipage}
    \hspace{0.02\textwidth}
  \begin{minipage}{0.40\textwidth}
    \begin{overpic}[width=\linewidth]{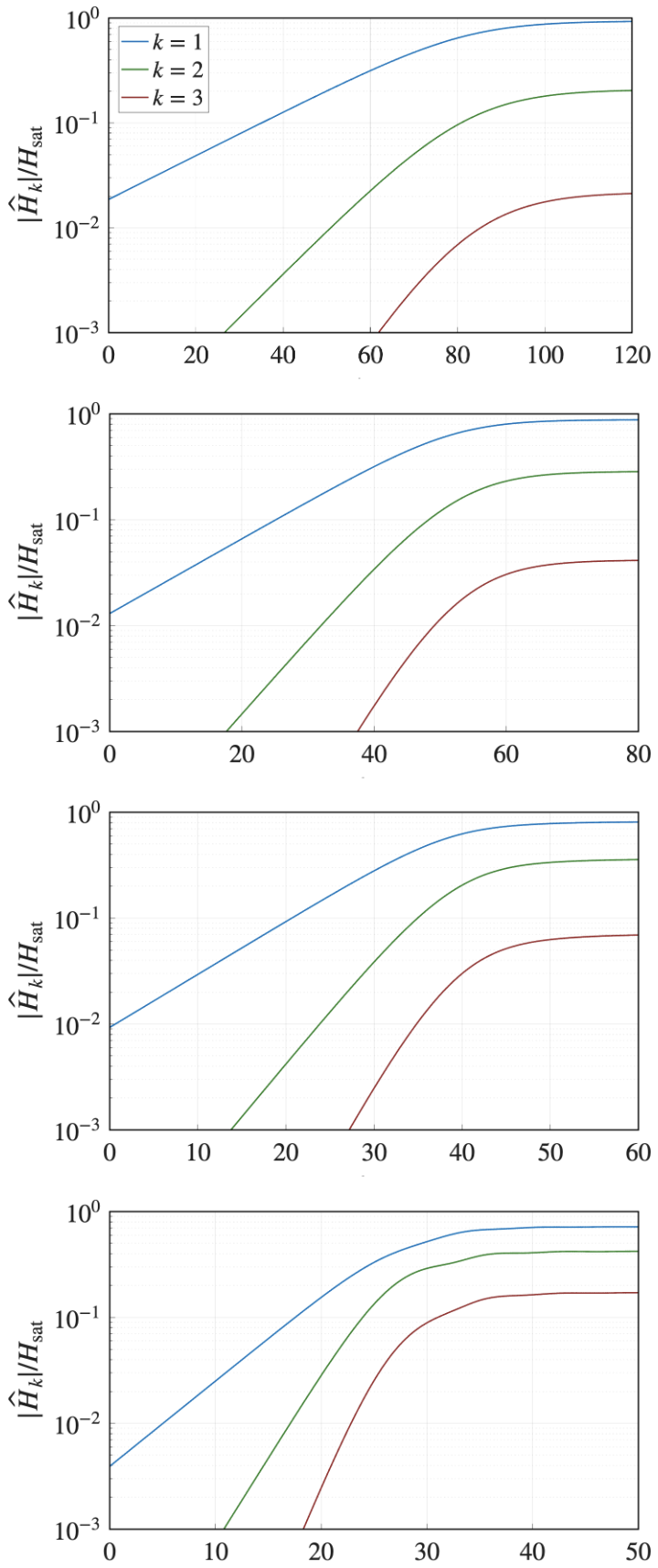}
      \put(0,98){(\textit{b})}
    \end{overpic}
    \vspace{0.3em}\makebox[\linewidth]{\fontsize{10}{12}\selectfont $t/T_{\mathrm{sat}}$}

  \end{minipage}

  \caption{Harmonic amplitudes for $U/U_L= 1.13,1.33,1.58, 2.25$.\quad (a) Experimental traces (from \citet{BartheletCharruFabre1995}, §§ 5.4 and 5.5);\quad (b) Model computation.}
  \label{fig:AoH}
\end{figure}

We proceed with the evolution of the first three harmonics of the interfacial amplitude signal. Experimental results are given in the left column of figure \ref{fig:AoH}, with the corresponding model results in the right column. The agreement, beyond transients, is very good even for the higher values of $R$. More specifically, in figure~\ref{fig:AoH} the Reynolds number $R$ increases from top to bottom. Larger $R$ yields an earlier approach to saturation for every mode $k$. The experimental traces display small non-monotone wiggles or brief dips during transients, which probably arise from random initial conditions in the laboratory and are not reproducible from run to run. The evolution time and the amplitude of harmonics are slightly different from the experiment up to a scaling. Despite small differences, the trends align closely: the ordering of appearance of higher modes due to nonlinear interactions, the systematic reduction of time taken to reach steady states as $R$ increases, and the progressive narrowing of the difference between the plateau levels of $k=1, 2, 3$ as $R$ increases. We can conclude that the model performs remarkably in reproducing the experiments.

\section{Conclusions}\label{sec:discussion}

We extended the work of \citet{KalogirouCimpeanuKeavenyPapageorgiou2016R1} to provide a comprehensive comparison between the nonlocal model and the experiments. We performed a more targeted sweep over $R$ and $\Lambda$, focusing on  bistability phenomena reported in experiments. In all cases studied we found remarkable agreement with the experiments. This can be attributed to the nonlocal terms in the model that capture inertial effects in the thick layer. Moreover, we identify and study in detail a new symmetry-breaking travelling-wave branch (branch~2$^\ast$) bifurcating from the $\pi$-periodic bimodal branch. We also map the broader travelling-wave structure, including higher-wavenumber branches (branches~3 and~4), and report time-periodic attractors arising when these travelling waves lose stability. These features were not reported previously to our knowledge.


\citet{BartheletCharruFabre1995} reported bistability for $U > 2.4U_L$, theoretically we find bistability also at $U / U_L = 1.06, 1.58, 3.50$ ($R = 309, 461, 1021$), with the corresponding bistability regions being $\Lambda \in [0.023, 0.041]$, $\Lambda \in [0.019, 0.042]$, and $\Lambda \in [0.014, 0.031]$. Including the bistability case of the experiment, we observe that as $R$ increases, branch 2 bifurcates from the trivial solution, becomes stable at smaller $\Lambda$, and finally loses stability earlier. For even larger $R$, the bistability window narrows. The range $R \in [400, 800]$ provides a relatively wide window in which two distinct saturated waves are observable. Nevertheless, the basins of attraction vary with $R$, so it is not surprising that some bistable regimes were not observed experimentally. 

Our results provide a detailed guide for parameter choices in future experiments focused on bistability. Indeed, the model suggests that bistability may be more common than previously thought: beyond the classical branch~1/branch~2 coexistence, we find additional multistability involving the symmetry-broken branch~2$^\ast$ as well as time-periodic long-wave states. A natural next step is to analytically prove the symmetry-breaking (pitchfork) bifurcation that produces branch~2$^\ast$ from branch~2 at $\Lambda \approx 0.0358$, and to test experimentally the predicted branch~2/branch~2$^\ast$ bistability for $\Lambda \in [0.057, 0.090]$. The higher-wavenumber travelling-wave families (branches~3 and~4) and the periodic orbits reported here further motivate experimental and numerical studies of multimodal and time-dependent interfacial long-wave dynamics in two-layer Couette flow.

\vspace{.2cm}

\noindent{\bf{Funding.}} The work of PG was supported by a Wolfson fellowship of the Royal Society. The work of DTP was partly supported by CBET-EPSRC grant EP/V062298/1. 

\noindent{\bf{Declaration of interests.}} The authors report no conflict of interest.

\bibliographystyle{jfm}
\bibliography{jfm}

@Article{Yih1967,
  author  = {C.-S. Yih},
  title   = {Instability due to viscosity stratification},
  journal = {J. Fluid Mech.},
  year    = {1967},
  volume  = {27},
  pages   = {337--352}
}

@Article{HooperBoyd1983,
  author  = {A. P. Hooper and W. G. C. Boyd},
  title   = {Shear-flow instability at the interface between two viscous fluids},
  journal = {J. Fluid Mech.},
  year    = {1983},
  volume  = {128},
  pages   = {507--528}
}

@Article{Hooper1985,
  author  = {A. P. Hooper},
  title   = {Long-wave instability at the interface between two viscous fluids: thin-layer effects},
  journal = {Phys. Fluids},
  year    = {1985},
  volume  = {28},
  pages   = {1613--1618}
}

@Article{Kalogirou2018,
  author  = {A. Kalogirou},
  title   = {Instability of two-layer film flows due to the interacting effects of surfactants, inertia, and gravity},
  journal = {Phys. Fluids},
  year    = {2018},
  volume  = {30},
  pages   = {030707}
}

@Article{Renardy1985,
  author  = {Y. Y. Renardy},
  title   = {Instability at the interface between two shearing fluids in a channel},
  journal = {Phys. Fluids},
  year    = {1985},
  volume  = {28},
  pages   = {3441--3443}
}

@Article{Renardy1987,
  author  = {Y. Y. Renardy},
  title   = {The thin-layer effect and interfacial stability in a two-layer Couette flow with similar liquids},
  journal = {Phys. Fluids},
  year    = {1987},
  volume  = {30},
  pages   = {1627--1637}
}

@Article{HooperGrimshaw1985,
  author  = {A. P. Hooper and R. Grimshaw},
  title   = {Nonlinear instability at the interface between two viscous fluids},
  journal = {Phys. Fluids},
  year    = {1985},
  volume  = {28},
  pages   = {37--45}
}

@Article{HooperGrimshaw1988,
  author  = {A. P. Hooper and R. Grimshaw},
  title   = {Travelling wave solutions to the Kuramoto--Sivashinsky equation},
  journal = {Wave Motion},
  year    = {1988},
  volume  = {10},
  pages   = {405--420}
}

@Article{CharruFabre1994,
  author  = {F. Charru and J. Fabre},
  title   = {Long waves at the interface between two viscous fluids},
  journal = {Phys. Fluids},
  year    = {1994},
  volume  = {6},
  number  = {3},
  pages   = {1223--1235}
}

@Article{BartheletCharruFabre1995,
  author  = {P. Barthelet and F. Charru and J. Fabre},
  title   = {Experimental study of interfacial long waves in a two-layer shear flow},
  journal = {J. Fluid Mech.},
  year    = {1995},
  volume  = {303},
  pages   = {23--53}
}

@Article{KalogirouCimpeanuKeavenyPapageorgiou2016R1,
  author  = {A. Kalogirou and R. C{\^\i}mpeanu and E. E. Keaveny and D. T. Papageorgiou},
  title   = {Capturing nonlinear dynamics of two-fluid Couette flows with asymptotic models},
  journal = {J. Fluid Mech.},
  year    = {2016},
  volume  = {806},
  pages   = {R1},
  doi     = {10.1017/jfm.2016.612}
}

@Article{PapageorgiouTanveer2019,
  author  = {D. T. Papageorgiou and S. Tanveer},
  title   = {Analysis and computations of a non-local thin-film model for two-fluid shear driven flows},
  journal = {Proc. R. Soc. A},
  year    = {2019},
  volume  = {475},
  number  = {2230},
  pages   = {20190367},
  doi     = {10.1098/rspa.2019.0367}
}

@Article{KatsiavriaPapageorgiou2022,
  author  = {A. Katsiavria and D. T. Papageorgiou},
  title   = {Nonlinear waves in viscous multilayer shear flows in the presence of interfacial slip},
  journal = {Wave Motion},
  year    = {2022},
  volume  = {114},
  pages   = {103018},
  doi     = {10.1016/j.wavemoti.2022.103018}
}

@Article{PapageorgiouTanveer2023,
  author  = {D. T. Papageorgiou and S. Tanveer},
  title   = {Singular effect of interfacial slip for an otherwise stable two-layer shear flow: analysis and computations},
  journal = {Proc. R. Soc. A},
  year    = {2023},
  volume  = {479},
  number  = {2272},
  pages   = {20220720},
  doi     = {10.1098/rspa.2022.0720}
}

@article{KalogirouPapageorgiou2016,
  author  = {Kalogirou, A. and Papageorgiou, D. T.},
  title   = {Nonlinear dynamics of surfactant-laden two-fluid Couette flows in the presence of inertia},
  journal = {J. Fluid Mech.},
  volume  = {802},
  pages   = {5--36},
  year    = {2016},
  doi     = {10.1017/jfm.2016.429},
}

@article{Kalogirou-et-al-2012,
  author  = {Kalogirou, A. and Papageorgiou, D. T. and Smyrlis Y.-S.},
  title   = {Surfactant destabilization and non-linear phenomena in two-fluid shear flows at small Reynolds numbers},
  journal = {IMA J. Appl. Math.},
  volume  = {77},
  pages   = {351--360},
  year    = {2012},
  doi     = {10.1093/imamat/hxs035},
}

@article{Akrivis-et-al-IMAJNA,
  author  = {Akrivis, G. and Papageorgiou, D. T. and Smyrlis Y.-S.},
  title   = {Linearly implicit methods for a semilinear parabolic system arising in two-phase flows},
  journal = {IMA J. Num. Anal.},
  volume  = {31},
  pages   = {299--321},
  year    = {2011},
  doi     = {10.1093/imanum/drp034},
}

@article{Kalogirou_Cimpeanu_Blyth_2020,
  author  = {Kalogirou, A. and C{\^i}mpeanu, R. and Blyth, M. G.},
  title   = {Asymptotic modelling and direct numerical simulations of multilayer pressure-driven flows},
  journal = {Eur. J. Mech. B/Fluids},
  volume  = {80},
  pages   = {195--205},
  year    = {2020}
}

@article{Kalogirou_Blyth_2023,
  author  = {Kalogirou, A. and Blyth, M. G.},
  title   = {Nonlinear dynamics of unstably stratified two-layer shear flow in a horizontal channel},
  journal = {J. Fluid Mech.},
  volume  = {955},
  pages   = {A32},
  year    = {2023},
  doi     = {10.1017/jfm.2022.1070}
}

@article{Katsiavria_Papageorgiou_2024,
  author  = {Katsiavria, A. and Papageorgiou, D. T.},
  title   = {Stability analysis of viscous multi-layer shear flows with interfacial slip},
  journal = {IMA J. Appl. Math.},
  volume  = {89},
  number  = {2},
  pages   = {279--325},
  year    = {2024}
}

\end{document}